\documentclass[11pt]{article}

\usepackage{bm, commath}
\usepackage{natbib}
\usepackage{caption}
\usepackage{graphicx}
\usepackage{subcaption}
\usepackage{amsmath, amsfonts, amsthm}
\usepackage{float}
\usepackage{booktabs,siunitx}
\usepackage{url}
\usepackage{multirow}
\usepackage{amssymb}
\usepackage{bm,bbm}
\usepackage{mathrsfs}
\usepackage{multirow}
\usepackage{authblk}

\usepackage{listings}
\allowdisplaybreaks
\usepackage{bbm}
\usepackage{textcomp}
\usepackage[margin=1in]{geometry}
\usepackage{authblk}
\usepackage[ruled]{algorithm2e}
\SetKwInput{KwParam}{Parameter}
\SetAlgoCaptionLayout{centerline}

\usepackage{sectsty}
\usepackage[onehalfspacing]{setspace}
\usepackage[usenames,dvipsnames]{color}
\usepackage{hyperref}
\usepackage[utf8]{inputenc}
\usepackage{float}
\usepackage{tikz}
\usetikzlibrary{trees}
\usetikzlibrary{shapes,decorations,arrows,calc,arrows.meta,fit,positioning}
\usetikzlibrary{shapes.multipart}
\tikzset{
    -Latex,auto,node distance =1 cm and 1 cm,semithick,
    state/.style ={ellipse, draw, minimum width = 0.7 cm},
    state1/.style ={ draw, minimum width = 0.7 cm},
    point/.style = {circle, draw, inner sep=0.04cm,fill,node contents={}},
    bidirected/.style={Latex-Latex,dashed},
    el/.style = {inner sep=2pt, align=left, sloped}
}

\newtheorem{theorem}{Theorem}

\newtheorem{corollary}{Corollary}

\newtheorem{condition}{Condition}
\newtheorem{lemma}{Lemma}

\usepackage{mathtools}

\usepackage{xcolor}
\usepackage{xr}
\makeatletter
\renewcommand{\algocf@captiontext}[2]{#1\algocf@typo. \AlCapFnt{}#2} % text of caption
\def\@algocf@capt@plain{top}
\renewcommand{\algocf@makecaption}[2]{%
  \addtolength{\hsize}{\algomargin}%
  \sbox\@tempboxa{\algocf@captiontext{#1}{#2}}%
  \ifdim\wd\@tempboxa >\hsize%     % if caption is longer than a line
  \hskip .5\algomargin%
  \parbox[t]{\hsize}{\algocf@captiontext{#1}{#2}}% then caption is not centered
  \else%
  \global\@minipagefalse%
  \hbox to\hsize{\box\@tempboxa}% else caption is centered
  \fi%
  \addtolength{\hsize}{-\algomargin}%
}
\makeatother

\def\E{\mathbbm{E}}
\def\R{\mathbbm{R}}
\def\bz{\bm{z}}
\def\bZ{\bm{Z}}

\def\gam{\gamma}
\def\eps{\epsilon}

\def\sig{\sigma}
\def\P{\mathbbm{P}}
\def\V{\mathcal{V}}

\def\F{\mathcal{F}}
\def\H{\mathcal{H}}
\def\Vdy{\mathcal{V}_{D\rightarrow Y}}
\def\Vyd{\mathcal{V}_{Y\rightarrow D}}
\def\bgam{\bm{\gamma}}
\def\bpi{\bm{\pi}}

\def\betayd{\beta_{Y \rightarrow D}}
\def\betady{\beta_{D \rightarrow Y}}
\def\bgam{\bm{\gamma}}
\def\bpi{\bm{\pi}}
\def\dy{D\rightarrow Y}
\def\yd{Y\rightarrow D}
\def\Gam{\Gamma}

\begin{document}

\sectionfont{\bfseries\large\sffamily}%

\subsectionfont{\bfseries\sffamily\normalsize}%

\title{
%A Focusing Framework for Bi-Directional Mendelian Randomization\\
A Focusing Framework for Testing Bi-Directional Causal Effects with GWAS Summary Data 
}

\author{Sai Li}
\affil{Institute of Statistics and Big Data, Renmin University of China, China.\thanks{saili@ruc.edu.cn}}

\author{Ting Ye}
\affil{Department of Biostatistics, University of Washington,  Seattle, Washington, U.S.A.\thanks{tingye1@uw.edu}}

\maketitle

\vspace{0.5 cm}
\noindent
\begin{abstract}
	 Mendelian randomization (MR) is a powerful method that uses genetic variants as instrumental variables (IVs) to infer the causal effect of a modifiable exposure on an outcome. Although recent years have seen many extensions of basic MR methods to be robust to certain violations of assumptions, few methods were proposed to infer bi-directional causal relationships, especially for phenotypes with limited biological understandings. The presence of horizontal pleiotropy adds another layer of complexity. In this article, we show that assumptions for common MR methods are often impossible or too stringent in the existence of bi-directional relationships. We then propose a new focusing framework for testing bi-directional causal effects between two traits with possibly pleiotropic genetic variants. Our proposal can be coupled with many state-of-art MR methods. We provide theoretical guarantees on the Type I error and power of the proposed methods. We demonstrate the robustness of the proposed methods using several simulated and real datasets.
\end{abstract}


\section{Introduction}

%{\color{red}	Instrumental variables (IVs) are widely used to learn causal relationships in existence of unmeasured confounders. [TY: this sentence feels a little redundant, as unmeasured confounders also appear in the next sentence.]}

	Mendelian randomization (MR) leverages genetic variation to infer the causal effect of a modifiable exposure on an outcome in the presence of unmeasured confounding \citep{Davey03, sanderson2022mendelian}. Because of the increasing availability of genetic data and a growing set of statistical methods,  the number and range of MR studies have expanded rapidly in the past decade	\citep{markozannes2022systematic}. When performed rigorously, MR studies can provide important insights into the pathogenic mechanism of diseases \citep{Holmes:2017, Pingault:2018aa, Adam:2019aa}.

%Why bidirectional causal relationship is of interested.
Most existing MR studies assume a putative one-directional causal relationship between two traits. %That is, there is no causal effect of the specified outcome on the specified exposure. 
However, this assumption can be restrictive in some practical scenarios.
First, a bi-directional causal relationship may exist in many applications \citep{DaveySmith14}.   For instance, \cite{carreras2018role} finds evidence that smoking reduces BMI  and higher BMI increases the risk of smoking;  \citet{carrasquilla2021mendelian} finds that long sedentary time can increase BMI and higher BMI can increase the sedentary time. Furthermore, even if a one-directional causal relationship is plausible, the causal direction between two traits can be unknown a priori. Hence, a model accounting for bi-directional causal effects is needed to infer the causal directions between a pair of traits.

Learning bi-directional causal relationships has two major challenges. 
First, the statistical essence of MR is using genetic variants (Single Nucleotide Polymorphisms, shorthanded as SNPs) that are associated with the exposure but have no direct effect on the outcome as instrumental variables (IVs). However, in presence of a bi-directional casual relationship, it can be difficult to distinguish between a SNP having its primary influence on the  exposure or the outcome, because SNPs appearing to be associated with the exposure can have their primary influence on either variable \citep{DaveySmith14}. To avoid this limitation,    it is recommended by \cite{DaveySmith14} to utilize SNPs with known functionality, which unfortunately is not always feasible for many traits of interest. The second challenge is the horizontal pleiotropy \citep{Verbanck:2018aa}, which occurs when  SNPs influence the outcome through pathways other than the given exposure. This violates one of the valid IV assumptions, the exclusion restriction condition, which requires that the SNPs affect the outcome exclusively through the exposure \citep{Burgess:2015ab}. Although many statistical methods have been developed recently to weaken the exclusion restriction condition, few methods allow for the existence of a bi-directional causal relationship.

\subsection{Prior works}
\label{subsec: prior works}

Inferring bi-directional causal relationships between two traits has received increased attention in various applications recently. The common practice is to conduct an MR analysis for each direction, but it lacks statistical guarantees and, according to our results in Section \ref{sec2}, can be questionable if SNPs used for two directions are not selected carefully. For these reasons, a recent primer by \cite{sanderson2022mendelian} cautions against this practice. \citet{richmond2017investigating} considers estimating the bi-directional relationship between insulin and increased adiposity by using two independent sets of SNPs with well-understood functionality as two sets of valid IVs, one for each trait, and performing MR analyses in both directions. However, the validity of this method is not proved and, as the authors commented, it can be hard in practice to find two independent sets of valid IVs. \citet{darrous2020simultaneous} proposes a latent heritable confounder MR method	assuming a hierarchical prior on the true parameters to  estimate the  bi-directional causal effects for two traits,  but the method is sensitive to prior misspecification and computationally unstable. Assuming the causal effect is only one-directional,  \cite{hemani2017orienting} proposes a Steiger's test with a single valid IV to infer the causal direction, which is later extended by \citet{xue2020inferring} to allow for multiple valid IVs and horizontal pleiotropy.

Much progress has been made to deal with horizontal pleiotropy when assuming the causal effect is one-directional,  which mostly falls into two categories. The first category assumes that pleiotropy only involves a small  proportion of SNPs. For example, \citet{Han:2008aa}, \citet{Bowden16}, \citet{Kang16}, and \citet{Wind19} assume that  at least 50\% of the IVs are valid.  \citet{Bowden17}, \citet{TSHT}, \cite{Wind21}, and \citet{guo2021post} impose the plurality rule, or the zero-mode assumption, which assumes that the number of valid IVs is larger than any number of invalid IVs with the same ratio estimator limit. The second category imposes some structure on the  pleiotropic effects. Along this line,  \citet{Bowden15} and \citet{Kole15} propose the InSIDE assumption, i.e., the SNPs'
 pleiotropic effects on the outcome are uncorrelated with their effects on the exposure.  \cite{Zhao:2019aa,zhao2018statistical} and \cite{ye2021debiased} additionally assume that  the pleiotropic effects have zero mean, which is known as balanced horizontal pleiotropy.  Some other methods include  \cite{Tchetgen2019_GENIUS, ye2021genius, Sun2019, qi2019mendelian}, and \cite{morrison2020ng}. In contrast,  in the presence of bi-directional relationships, few works have formally studied how to deal with pleiotropic effects except for \cite{darrous2020simultaneous}.

\subsection{Our contributions}

	Our contributions are two-folded. First, we formally evaluate common identifiability assumptions  in MR under a bi-directional causal model. We show that existing identification conditions on the proportion of invalid IVs, such as the valid rule, majority rule, and  plurality rule, cannot hold  for both directions simultaneously. The InSIDE assumption and balanced horizontal pleiotropy assumption hold only under a peculiar scenario and lacks scientific underpinning under bi-directional relationships. These facts caution the attempt to directly apply state-of-art one-directional MR methods to both directions.

Second,  we propose a new  \emph{focusing} framework for inferring bi-directional causal effects between two traits with possibly invalid IVs. Our proposal consists of two steps. Under the null hypothesis that one trait has no effect on the other trait, we develop a strategy to select valid IVs for this causal direction. We term the set of  selected variants as the ``focused set''. Then we test the null hypothesis by
applying a one-directional MR method using variants in the focused set with appropriate adjustment for post-selection. As such,  many state-of-art MR methods can be coupled with the focusing framework.
We apply these two steps for both directions. In Section \ref{sec3}, we prove the Type I error control of the proposed focused testing methods and establish their power performance.

We apply the proposed focusing framework coupled with the inverse-variance weighted estimator and median estimator  to interrogate the pairwise causal relationship among 5 phenotypes. Among the 10 pairs of traits considered, 
our proposed methods detected four one-directional plus one bi-directional causal effects that are supported by the literature. In contrast, conducting MR analyses for both directions (using either the  inverse-variance weighted or MR-Median estimator), each using variants strongly associated with the putative exposure, leads to many implausible discoveries.

\subsection{Organization}
The rest of the paper is organized as follows. Section \ref{sec2} introduces notation and assumptions, and evaluate common MR assumptions  under a bi-directional causal model. Section \ref{sec3}  proposes our focusing framework and establishes the statistical properties. Results from simulations and real applications are presented in Section \ref{sec-simu} and \ref{sec-data}, respectively. We conclude with a summary and discussion in Section \ref{sec-diss}. Technical proofs and further numerical results are in the supplementary material.

\section{Identification in Bi-directional Causal Models}
\label{sec2}
In this section, we introduce the bi-directional causal model. Then we define the valid and invalid IVs and show how they affect the identification of causal effects. We also evaluate the validity of some commonly-imposed IV assumptions in existence of bi-directional causal relationships.

\subsection{Bi-directional causal models}
Let $\bZ\in\R^p$ be a column vector of $p$ SNPs, which are possibly invalid IVs, and $D\in\R$ and $Y\in\R$ a pair of phenotypes. We start with the following structural equations: 
 \begin{align}
 \label{eq-bi-MR}
 \left\{
 \begin{array}{ll}
	Y=\mu_Y+D\betady+\bZ^T\bpi_Y+ u_Y,  ~\E[ u_Y |\bZ]=0\\
	D=\mu_D+Y\betayd+\bZ^T\bpi_D+ u_D, ~\E[u_D|\bZ]=0,
	\end{array}
	\right.
\end{align}
where $\mu_Y$ and $\mu_D$ are the intercepts, $\betady$ is the causal effect of $D$ on $Y$ and $\betayd$ is the causal effect of $Y$ on $D$, $\bpi_Y= (\pi_{Y,1}, \dots, \pi_{Y,p})^T $ and $\bpi_D =  (\pi_{D,1}, \dots, \pi_{D,p})^T$ respectively denote the direct effects of SNPs $\bZ$ on $Y$ and $D$, and $u_Y, u_D$ are errors that are correlated due to the unmeasured confounders $U$. The structural equations in (\ref{eq-bi-MR}) describe the bi-directional causal relationship in its equilibrium state. Models of this kind are termed simultaneous equation models in econometrics  \citep{Hausman1983specification}.  If $\betayd=0$, then equations in (\ref{eq-bi-MR}) reduce to the commonly-assumed one-directional  structural equations with the exposure $D$ and outcome $Y$.

Let $\bm{B}=\begin{pmatrix} 1 &-\betayd\\
-\betady & 1\end{pmatrix}$.  Following \citet{Hausman1983specification}, we assume that $\bm{B}$ is a non-singular matrix, i.e., $\betady\betayd\neq 1$. Then it is easy to derive 
\begin{align}
\label{eq-data}
(Y,D)=(\mu_Y,\mu_D)\bm{B}^{-1}+\bZ^T\begin{pmatrix}
\bpi_Y &\bpi_D
\end{pmatrix}\bm{B}^{-1}+(u_Y ,u_D)\bm{B}^{-1},
\end{align}
which gives the following reduced-form equations for the phenotypes 
\begin{align}
&\E[Y|\bZ=\bz]=\tilde{\mu}_Y +\bz^T\bgam_Y~~\text{and}~~\E[D|\bZ=\bz]=\tilde{\mu}_D+\bz^T\bgam_D, ~\text{where}\label{eq-out}\\
&\bgam_Y=\frac{\bpi_Y+\bpi_D\betady}{1-\betady\betayd}~~\text{and}~~\bgam_D=\frac{\bpi_D+\bpi_Y\betayd}{1-\betady\betayd},\label{eq-gam}
\end{align}
 and $(\tilde{\mu}_Y, \tilde{\mu}_D)=(\mu_Y,\mu_D)\bm{B}^{-1}$ are the intercepts in the reduced-form equations.
 Notice that 
 $\bgam_Y= (\gamma_{Y,1}, \dots, \gamma_{Y,p})^T$ and $\bgam_D=(\gamma_{D,1}, \dots, \gamma_{D,p})^T$, both linear combinations of $(\bpi_D,\bpi_Y)$,  are identifiable as long as $\E[\bZ\bZ^T]$ is positive definite. However,  without further conditions, the causal effects $\betayd$ and $ \betady$ are not identifiable because there are $2p+2$ unknown parameters but only $2p$ constraints in (\ref{eq-gam}).

\subsection{Three types of IVs in the bi-directional models}
\label{sec2-2}

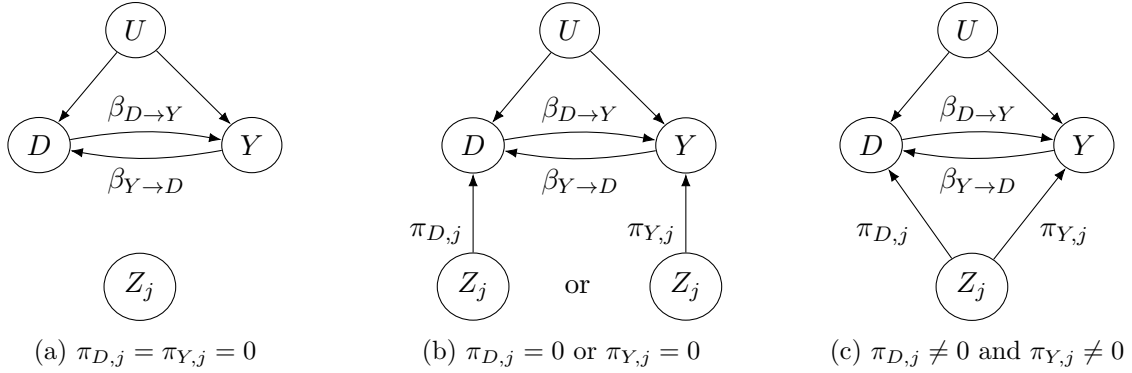
\begin{figure}[t]
	\centering
	\begin{subfigure}[b]{0.33\textwidth}
		\centering
		\begin{tikzpicture}
			% x node set with absolute coordinates
			\node[state] (x) at (0,0) {$D$};
			
			% y node set relative to x.
			% Locations can be:
			% right,left,above,below,
			% above left,below right, etc
			\node[state] (y) [right =of x, xshift=1cm] {$Y$};
			
			\node[state] (u) [above right =of x,xshift=-0.3cm,yshift=0cm] {$U$};
			
			\node[state] (z) [below right =of x,xshift=-0.3cm,yshift=-0.3cm] {$Z_j$};
			% Directed edge
			\path (x) edge[bend left=10]  node[above]{$\betady$} (y);
			\path (y) edge [bend right=-10] node[below]{$\betayd$} (x);
			\path (u) edge node[above]{} (x);
			\path (u) edge node[above]{} (y);
			%\path (z) edge node[below left]{$\pi_D$} (x);
			%\path (x) edge node[above]{} (z);
			%\path (z) edge node[below right]{$\pi_Y$} (y);
			%\path (y) edge node[above]{} (z);
		\end{tikzpicture} 
		\caption{$ \pi_{D,j} = \pi_{Y,j} = 0 $}
	\end{subfigure}\hfill
	\begin{subfigure}[b]{0.33\textwidth}
		\centering
		\begin{tikzpicture}
			% x node set with absolute coordinates
			\node[state] (x) at (0,0) {$D$};
			
			% y node set relative to x.
			% Locations can be:
			% right,left,above,below,
			% above left,below right, etc
			\node[state] (y) [right =of x, xshift=1cm] {$Y$};
			
			\node[state] (u) [above right =of x,xshift=-0.3cm,yshift=0cm] {$U$};
			
			\node[state] (zx) [below =of x,yshift=-0.05cm] {$Z_j$};
			\node[state] (zy) [below =of y,yshift=-0.05cm] {$Z_j$};
			\node (or) [right= of zx, xshift=-0.4cm] {or};
			% Directed edge
			\path (x) edge[bend left=10]  node[above]{$\betady$} (y);
			\path (y) edge [bend right=-10] node[below]{$\betayd$} (x);
			\path (u) edge node[above]{} (x);
			\path (u) edge node[above]{} (y);
			\path (zx) edge node[below left]{$\pi_{D,j}$} (x);
			\path (zy) edge node[below left]{$\pi_{Y,j}$} (y);
			%\path (x) edge node[above]{} (z);
			%\path (z) edge node[below right]{$\pi_Y$} (y);
			%\path (y) edge node[above]{} (z);
		\end{tikzpicture} 
		\caption{$ \pi_{D,j} = 0 $ or  $\pi_{Y,j} = 0 $}
	\end{subfigure}\hfill
	\begin{subfigure}[b]{0.33\textwidth}
		\centering
		\begin{tikzpicture}
			% x node set with absolute coordinates
			\node[state] (x) at (0,0) {$D$};
			
			% y node set relative to x.
			% Locations can be:
			% right,left,above,below,
			% above left,below right, etc
			\node[state] (y) [right =of x, xshift=1cm] {$Y$};
			
			\node[state] (u) [above right =of x,xshift=-0.3cm,yshift=0cm] {$U$};
			
			\node[state] (z) [below right =of x,xshift=-0.3cm,yshift=-0.3cm] {$Z_j$};
			% Directed edge
			\path (x) edge[bend left=10]  node[above]{$\betady$} (y);
			\path (y) edge [bend right=-10] node[below]{$\betayd$} (x);
			\path (u) edge node[above]{} (x);
			\path (u) edge node[above]{} (y);
			\path (z) edge node[below left]{$\pi_{D,j}$} (x);
			%\path (x) edge node[above]{} (z);
			\path (z) edge node[below right]{$\pi_{Y,j}$} (y);
			%\path (y) edge node[above]{} (z);
		\end{tikzpicture} 
		\caption{$ \pi_{D,j}\neq 0 $ and $ \pi_{Y,j}\neq 0 $}
	\end{subfigure}\hfill
	\caption{Illustration of three types of IVs: null IV (left), valid IV for each direction (middle), and pleiotropic IV (right).\label{DAG1}}
	\label{fig1}
\end{figure}

As illustrated in Figure \ref{DAG1}, we classify all the SNPs into three categories: (a) null IV: both $\pi_{D,j}$ and $\pi_{Y,j}$ are zero; (b) valid IV for one direction: one and only one of $\pi_{D,j}$ and $\pi_{Y,j}$ are nonzero; (c)  pleiotropic IV: both $\pi_{D,j}$ and $\pi_{Y,j}$ are nonzero. Our definition of valid IVs agrees with the conventional definition. Specifically, a valid IV for identifying $\betady$ is usually defined as having $\gam_{D,j}\neq 0, \pi_{Y,j}=0$  in the literature, which according to \eqref{eq-gam} is equivalent to our definition that $\pi_{D,j}\neq 0 ,\pi_{Y,j}=0$. Let $\V_{null}=\{j:\pi_{D,j}=\pi_{Y,j}=0\}$ denote the set of null IVs, $\mathcal{V}_{D\rightarrow Y}=\{j:\pi_{D,j}\neq 0 ,\pi_{Y,j}=0\}$ the set of valid IVs for identifying $\betady$, $\mathcal{V}_{Y\rightarrow D}=\{j:\pi_{Y,j}\neq 0 ,\pi_{D,j}=0\}$ the set of valid IVs for identifying  $\betayd$, and  $\V_{pl}=\{j:\pi_{D,j}\neq 0,\pi_{Y,j}\neq 0\}$ the set of pleiotropic IVs. These four sets $\V_{null}$, $\Vdy$, $\Vyd$, and $\V_{pl}$ provide a mutually exclusive and exhaustive partition of the set of candidate IVs. We also define the set of invalid IVs for direction $\dy$ as $\Vdy^c$, which includes null IVs, pleiotropic IVs, and valid IVs for the other direction.

By (\ref{eq-gam}), the null IVs have $\gam_{D,j}=\gam_{Y,j}=0$. Non-identification with null IVs is trivial. As  explained above, causal effects are also not identifiable based on the pleiotropic IVs because there are more parameters than the identification constraints. 
We now look into the one-directional valid IVs, i.e., those in $\Vdy$ or $\Vyd$, which play a key role incausal effect identification.

\begin{lemma}
\label{lem-b}
Suppose that $\betady\betayd\neq 1$ and \eqref{eq-gam} holds. For $j\in\Vdy$, 
\begin{align}
\label{eq2-b}
    \gam_{D,j}=\frac{\pi_{D,j}}{1-\betady\betayd}~~\text{and}~~\gam_{Y,j}=\frac{\pi_{D,j}\betady}{1-\betady\betayd}.
\end{align}
For $j\in\Vyd$,
\begin{align}
\label{eq1-b}
    \gam_{D,j}=\frac{\pi_{Y,j}\betayd}{1-\betady\betayd}~~\text{and}~~\gam_{Y,j}=\frac{\pi_{Y,j}}{1-\betady\betayd}.
\end{align}
As a consequence, $\{\gam_{Y,j}/\gam_{D,j}\}_{j\in \Vdy}=\betady$ and $\{\gam_{D,j}/\gam_{Y,j}\}_{j\in \Vyd}=\betayd$.
\end{lemma}

Based on the one-directional valid IVs, the ratios $\{\gam_{D,j}/\gam_{Y,j}\}_{j\in \Vyd}$ and $\{\gam_{Y,j}/\gam_{D,j}\}_{j\in \Vdy}$ identify $\betayd$ and $\betady$, respectively. Hence, knowing at least one valid IV for each direction is sufficient for identifying the pair of causal effects.

If $\Vyd$ and $\Vdy$ are unknown, bi-directional causal relationship can result in identifiability issues. This is because the ratios $\{\gam_{Y,j}/\gam_{D,j}\}_{j\in \Vdy\cup\Vyd}$ can have two modes: $\betady$ from $\Vyd$ and $1/\betayd$ from $\Vdy$, when $\betayd\neq 0$ and $\betady\neq 1/\betayd$.  This fact reveals that valid IVs for one direction are invalid IVs for the other direction. Hence, the issue of invalid IVs is naturally embedded in bi-directional causal models, even in the absence of pleiotropic IVs.

In addition, it is hard  to distinguish between $\Vdy$ and $\Vyd$ in a data-driven way without prior knowledge. If in the absence of reverse causation, i.e., $\betayd= 0$, the IVs  in $\Vyd$ are not associated with $D$ according to (\ref{eq1-b}) and thus can be easily removed based on the estimates of $\gamma_{D,j}$'s. With a bi-directional causal relationship, the IVs in  $\Vyd$ are associated with both $D$ and $Y$ as induced by the reverse causal effect; same for the IVs in $\Vdy$. This causes intrinsic difficulties in identifying bi-directional causal effects  
when we have little understanding of the SNPs' biological effects  \citep{DaveySmith14}, as formalized in Lemma \ref{lem-b2} below. 

\begin{lemma}
	\label{lem-b2}
	Suppose that $\betady\betayd\neq 1$ and \eqref{eq-gam} holds. Given $\{ \bgam_D, \bgam_Y \}_{j \in \Vdy\cup \Vyd} $ with unknown $\Vdy$ and $\Vyd$, neither $\betady$ nor $\betayd$ is identifiable.
\end{lemma}

%{\color{magenta}
%When we are interested in $\betady$, the reverse causal effect $\betayd\neq 0$ can induce pleiotropic effects on some IVs, the IVs in $\Vyd$.  
%From this perspective, the issues of invalid IVs are naturally embedded in the causal effect identification in bi-directional MR models, even in the absence of pleiotropic IVs, i.e., $\V_{pl}=\emptyset$.}

Before developing new methods to deal with this identification challenge, we first evaluate common identification conditions for MR under our bi-directional causal models (\ref{eq-bi-MR}).

\subsection{ Evaluation of common identification conditions for MR}
\label{sec3-2}
The  existing MR literature typically only considers a one-directional relationship (e.g., by assuming $\betayd=0$ in (\ref{eq-gam})), given the perception that MR is able to deal with reverse causation. This perception is certainly correct if we have valid IVs. However, if we only have a set of SNPs that are associated with the given exposure, it can be a mix of valid IVs for the exposure, valid IVs for the outcome (due to reverse causation), and pleiotropic IVs. Therefore, the commonly-imposed assumptions,  including the valid rule, majority rule, plurality rule, and InSIDE assumption, can be overly restrictive for studying the effect of the given exposure on the outcome. Moreover, some of these assumptions cannot hold simultaneously for both directions when bi-directional causal effects exist.

Define the set of relevant IVs for $D$ as $\mathcal{S}_D=\{j:\gam_{D,j}\neq 0\}$ and that for $Y$ as $\mathcal{S}_Y=\{j:\gam_{Y,j}\neq 0\}$. For the direction $\dy$, we say that the valid rule holds if all the relevant SNPs for $D$ are valid IVs for identifying  $\betady$, i.e., $\mathcal{S}_D=\Vdy\neq\emptyset$; analogously, the valid rule for $\yd$ says that $\mathcal{S}_Y=\Vyd\neq\emptyset$. The valid rule is the basis of the two-stage least squares and inverse-variance weighted estimators \citep{Burgess:2015ab}.

\begin{corollary}[Valid rule in the bi-directional model]
\label{cor-valid}
	Suppose that $\betady\betayd\neq 1$ and \eqref{eq-gam} holds.
(i) If $\betayd\neq 0$, then the valid rule for the direction $\dy$ holds only if $\Vyd$ is empty.
(ii) If $\betady\neq 0$ or $\betayd\neq 0$, then the valid rule cannot simultaneously hold for both directions.
\end{corollary}
Corollary \ref{cor-valid} is a direct consequence of Lemma \ref{lem-b}.
Part (i) implies that assuming the valid rule in one direction implicitly assumes that
there is no valid IV for the other direction. Part (ii) implies that one cannot simultaneously assume the valid rule for both directions given a set of candidate IVs $\{Z_1,\dots, Z_p\}$.
%This is because, $S_D$ is indeed the union of $\Vdy$ and $\Vyd$ when $\betady\neq 0$ and $\betayd\neq 0$. By Lemma \ref{lem-b} these two sets will give different estimates of causal effects and cannot both be valid. This implies that assuming all the strong IVs are valid in each direction can be too strong to be true.}

Next, we look into a milder condition, the majority rule, in the bi-directional model. 
The majority rule for identifying $\betady$ requires that more than 50\% of the relevant SNPs are valid, i.e., $|\Vdy|>|\mathcal{S}_D|/2$; analogously, the majority rule for identifying $\betayd$ requires that $|\Vyd|>|\mathcal{S}_Y|/2$,   where $|A|$ denotes the cardinality of a set $A$. The majority rule is the basis of median-based methods  \citep{Bowden16,Kang16}. 
	
%	[YT: I made the change as we never use the term ``bi-directional majority rule'', but only refer to majority rule in either direction.]}

\begin{corollary}[Majority rule in the bi-directional model]
\label{cor-maj}
	Suppose that $\betady\betayd\neq 1$ and \eqref{eq-gam} holds. (i) If $\betayd\neq 0$, then the majority rule for the direction $\dy$ holds if and only if $|\Vyd|<|\mathcal{S}_D|/2$. (ii) If $\betady\neq 0$ and $\betayd\neq 0$, then the majority rule  for both directions cannot hold simultaneously. %Furthermore, if $\betady\betayd\neq 0$, then a necessary condition for the majority rule for direction $\betady$ is $|S_D\cap S_Y|>|S_D|/2$; a necessary condition for the majority rule for direction $\betayd$ is $|S_D\cap S_Y|>|S_Y|/2$.
\end{corollary}

Next, we investigate an even milder condition, the plurality rule. The plurality rule for identifying $\betady$ assumes that the mode of $\{\pi_{Y,j}/\gam_{D,j}\}_{j\in \mathcal{S}_D}$ is zero; analogously, the plurality rule for identifying $\betayd$ assumes that the mode of $\{\pi_{D,j}/\gam_{Y,j}\}_{j\in \mathcal{S}_Y}$ is zero. The plurality rule is the basis for mode-based methods \citep{Bowden17,TSHT}.

\begin{corollary}[Plurality rule in the bi-directional model]
\label{cor-plu}
	Suppose that $\betady\betayd\neq 1$ and \eqref{eq-gam} holds.  If $\betady\neq 0$ and  $\betayd\neq 0$, then the plurality rule for identifying $\betady$ and $\betayd$ cannot hold simultaneously. 
\end{corollary}

Corollaries \ref{cor-maj} and \ref{cor-plu} are consequences of Lemma \ref{lem-b}. These two corollaries imply that one cannot assume the majority rule or the plurality rule for both directions given a set of candidate IVs.

Finally, we look into another popular condition, the InSIDE assumption \citep{Bowden15}. The InSIDE
 assumption for the direction $\betady$ can be written as $(\bgam_D-\bar{\gam}_D)^T(\bpi_Y-\bar{\pi}_Y)= 0$ and that for the direction $\betayd$ can be written as $(\bgam_Y-\bar{\gam}_Y)^T(\bpi_D-\bar{\pi}_D)=0$, where $\bar{\gam}_D$, $\bar{\gam}_Y$, $\bar{\pi}_D$, $\bar{\pi}_Y$ are the averages of $\bgam_D$, $\bgam_Y$, $\bpi_D$, $\bpi_Y$, respectively. The InSIDE assumption is the basis of MR-Egger \citep{Bowden15,Kole15}. 

\begin{corollary}[InSIDE assumption in the bi-directional models]
\label{cor-inside}
	Suppose that $\betady\betayd\neq 1$ and \eqref{eq-gam} holds. The InSIDE assumption holds for the direction $\dy$ if and only if $\betayd=-(\bpi_Y-\bar{\pi}_Y)^T(\bpi_D-\bar{\pi}_D)/\|\bpi_Y-\bar{\pi}_Y\|_2^2$. 
The InSIDE assumption for the direction $\yd$ if and only if $\betady=-(\bpi_D-\bar{\pi}_D)^T(\bpi_Y-\bar{\pi}_Y)/\|\bpi_D-\bar{\pi}_D\|_2^2$.
\end{corollary}
 The key insight is that reverse causation can induce correlation between the SNPs' effects on exposure and their direct effects on the outcome, except under a peculiar scenario that the  reverse causation and the correlation between $\bpi_D$ and $\bpi_Y$ exactly cancels out. Hence, Corollary \ref{cor-inside} reveals that the InSIDE assumption and the stronger balanced horizontal pleiotropy assumption (discussed in Section \ref{subsec: prior works}) are overly restrictive and lack scientific underpinning   under bi-directional relationships.

As a summary, invalid IVs are intrinsic when considering bi-directional causal effects because valid IVs for one direction are invalid for the other direction. A bi-directional relationship also makes many established MR assumptions overly restrictive. Even more alarming is that, according to 	 Corollaries \ref{cor-valid}-\ref{cor-inside}, these MR assumptions cannot hold simultaneously for both directions, which prohibits directly applying state-of-art MR methods to both directions to infer bi-directional causal effects. Therefore, we need new methods to study bi-directional causal effects under milder assumptions.

\section{Testing Bi-Directional Causal Effects: A Focusing Framework}
\label{sec3}
 
In this section, we develop a new testing framework that can overcome the aforementioned challenges in inferring bi-directional causal effects and can be coupled with many state-of-art one-directional MR methods.

Under the structural equations in (\ref{eq-bi-MR}), the whole parameter space is $\{(\betady,\betayd):~\betady\betayd\neq 1\}$.
We consider three null subspaces (i) $\mathcal{H}^{(0,0)}:~\betady=\betayd=0$; (ii) $\mathcal{H}_{\dy}^{(0)}:~\betady=0$; (iii) $\mathcal{H}_{\yd}^{(0)}:~\betayd=0$. 
 Specifically, no causal effects exist in $\mathcal{H}^{(0,0)}$. No bi-directional causal effect exists in  $\mathcal{H}_{\dy}^{(0)}$ or $\mathcal{H}_{\yd}^{(0)}$. Testing for these three null hypotheses 	not only detects the existence of 
  	 bi-directional causal effects but also causal directions.
 
% These three null hypotheses distinguish the existence and directions of {\red causal effects}.

In the rest of this section, we present our results for testing $\mathcal{H}_{\dy}^{(0)}$. Methods and results for $\mathcal{H}_{\yd}^{(0)}$ can be obtained in the same way by switching $D$ and $Y$. The hypothesis $\mathcal{H}^{(0,0)}$ can be tested based on $\mathcal{H}_{\dy}^{(0)}$ and $\mathcal{H}_{\yd}^{(0)}$ with Bonferroni correction.  When jointly testing three hypotheses, the sequential method by \cite{rosenbaum2008testing} can be applied (described in the supplement).

\subsection{Key ideas}
\label{subsec: key ideas}

The main device of our proposal is that, for testing $\mathcal{H}_{\dy}^{(0)}$, we define a focused set $\widehat{\F}_{\dy}$ which can exclude most invalid IVs for direction $\dy$ under the null. Then we can apply established one-directional MR methods with SNPs in the focused set for causal inference. 

To be more specific, we focus on a widely-used setup in MR known as the two-sample summary-data MR, where two sets of summary statistics are obtained from two  genome-wide association studies (GWASs). The first set from one GWAS (sample size $n_D$) consists of $\hat{\gam}_{D,j}$, the estimated marginal association between the $j$th SNP and trait $D$, and its standard error $\xi_{D,j}$, $j=1,\dots, p$. The second set from another GWAS (sample size $n_Y$) consists of $\hat{\gam}_{Y,j}$, the estimated marginal association between the $j$-th SNP and trait $Y$, and its standard error $\xi_{Y,j}$, $j=1,\dots, p$. Hence, $\xi^2_{D,j}$ is of order $n_D^{-1}$ and $\xi_{Y,j}^2$ is of order $n_Y^{-1}$. The independence of $\hat{\gam}_{D,j}$ and $\hat{\gam}_{Y,j}$ is guaranteed by the two-sample MR design. As formally stated in Condition \ref{cond1}, we also assume that the $p$ SNPs in use are mutually independent after linkage disequilibrium pruning or clumping \citep{Hemani:2018ab}. 

We define the focused set for testing $\H_{\dy}^{(0)}$ as
\begin{align}
\label{eq-Fdy}
  \widehat{\F}_{\dy}=\left\{1\leq j\leq p: |\hat{\gam}_{Y,j}|\leq \xi_{Y,j}\tau_f,~|\hat{\gam}_{D,j}|\geq \xi_{D,j}\tau_s\right\},
\end{align}
where $\tau_f$ is the tuning parameter for excluding invalid IVs and $\tau_s$ is the tuning parameter for screening relevant SNPs. Setting $\tau_s> 0$ helps screen out null IVs as they can  diminish the testing power and it is widely adopted in practice. 

%Similarly,  for testing $\H_{\yd}^{(0)}$, we exchange the roles of $D$ and $Y$ and define
%\begin{align}
%\label{eq-Fyd}
%  \widehat{\F}_{\yd}=\left\{1\leq j\leq p: |\hat{\gam}_{D,j}|\leq \xi_{D,j}\tau_f,~|\hat{\gam}_{Y,j}|\geq \xi_{Y,j}\tau_s\right\}.
%\end{align}

To see why focusing works, notice that under $\H_{\dy}^{(0)}$,
\begin{align}
\label{gamY-margin}
\hat{\gam}_{Y,j}= \hat{\eps}_{Y,j}~\text{if}~j\in\Vdy\cup\V_{null};~~\hat{\gam}_{Y,j}=\pi_{Y,j}+\hat{\eps}_{Y,j}~\text{if}~j\in\Vyd\cup \V_{pl},
\end{align}
where $\hat{\eps}_{Y,j}$ denotes the noise part with mean zero. 
Because $\pi_{Y,j}\neq 0$ for $j\in\Vyd\cup \V_{pl}$, we know that $ \hat{\gam}_{Y,j}/\xi_{Y,j}$ from $\Vyd\cup \V_{pl}$ are further away from zero than those from $\Vdy\cup\V_{null}$. Therefore, restricting to a set of SNPs with relatively small $|\hat{\gam}_{Y,j}|/\xi_{Y,j}$ can screen out SNPs with nonzero direct effects on $Y$ and retain SNPs that are more likely to belong to $\Vdy\cup\V_{null}$. The choice of tuning parameters will be theoretically studied in the next subsection and numerically studied in Section \ref{sec-simu}.  

 %The same story holds for $\widehat{\F}_{\yd}$ under the null $\H_{\yd}^{(0)}$. 

Within the focused set, we can apply many state-of-art  one-directional MR methods with adjustment for post-selection. For example, one can calculate the inverse-variance weighted (IVW) estimator based on the focused set $\widehat{\F}_{\dy}$, which is 
\begin{align}
\label{psi-dy}
\hat{\psi}_{\dy}=\frac{\sum_{j\in\widehat{\F}_{\dy}}\hat{\omega}_{Y,j}\hat{\gam}_{Y,j}/\hat{\gam}_{D,j}}{\sum_{j\in\widehat{\F}_{\dy}}\hat{\omega}_{Y,j}}~\text{with}~~\hat{\omega}_{Y,j}=\frac{\hat{\gam}_{D,j}^2}{\xi_{Y,j}^2}.
\end{align} 
In (\ref{psi-dy}), $\hat{\gam}_{Y,j}/\hat{\gam}_{D,j}$ is the estimate of $\betady$ based on the $j$-th  SNP and can be unstable when the SNP is weak. The inverse-variance weighted (IVW)  estimator aggregates all the available estimates in the focused set and thus is more robust to weak IVs.  One can also apply the median estimator based on the focused set $\widehat{\F}_{\dy}$ 
\begin{align}
\label{psim-dy}
\hat{\psi}_{\dy}^{(m)}=\text{Median}\left(\left\{ \hat{\gam}_{Y,j}/\hat{\gam}_{D,j}\right\}_{j\in\widehat{\F}_{\dy}}\right).
\end{align} 
To be related to the existing methods, the conventional IVW and median estimators are ``\textit{overall}'' methods with $\tau_f=\infty$. Testing procedures based on the focused sets leverage the structure of the null hypothesis and remove most invalid IVs efficiently.
  
\begin{figure}[H]
\centering
\includegraphics[width=0.46\textwidth,height=5.2cm]{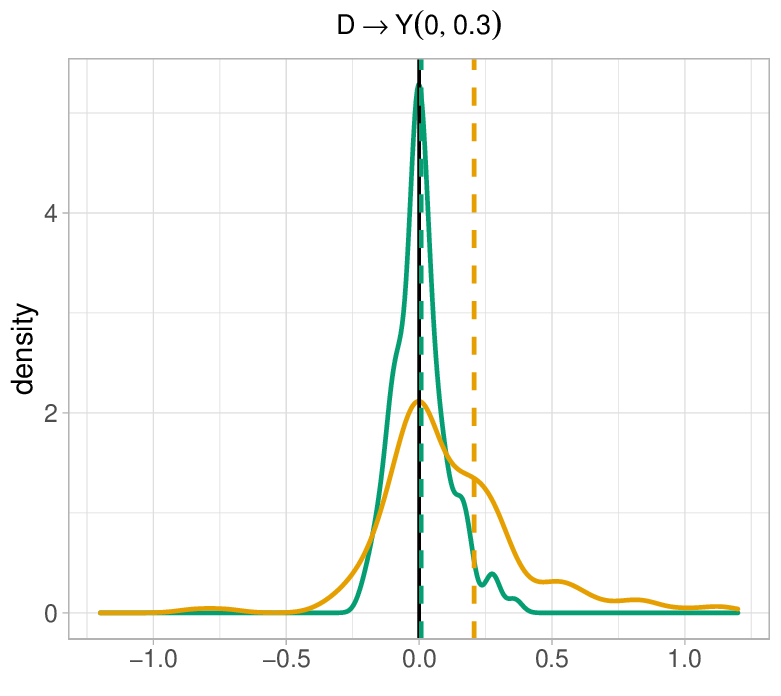}
\includegraphics[width=0.48\textwidth,height=5.2cm]{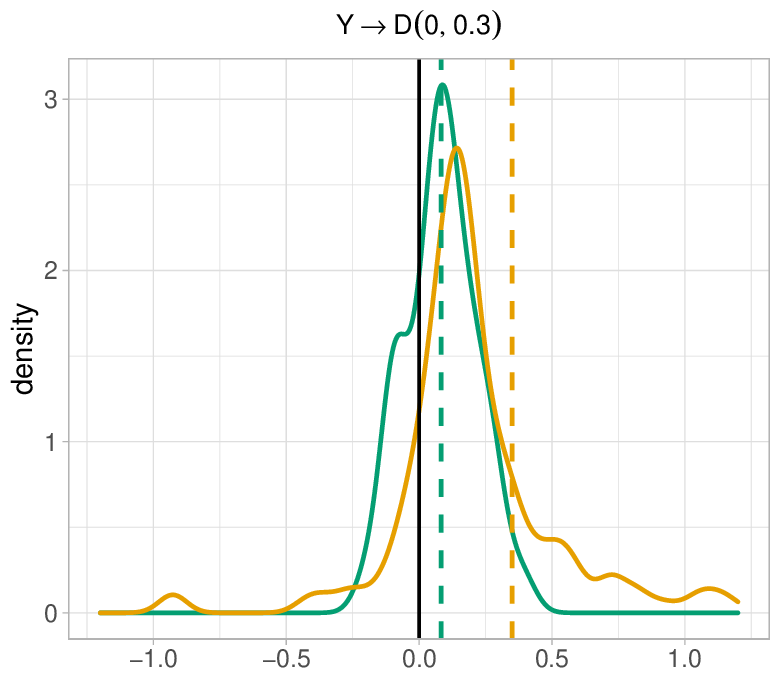}
\includegraphics[width=0.46\textwidth,height=5.2cm]{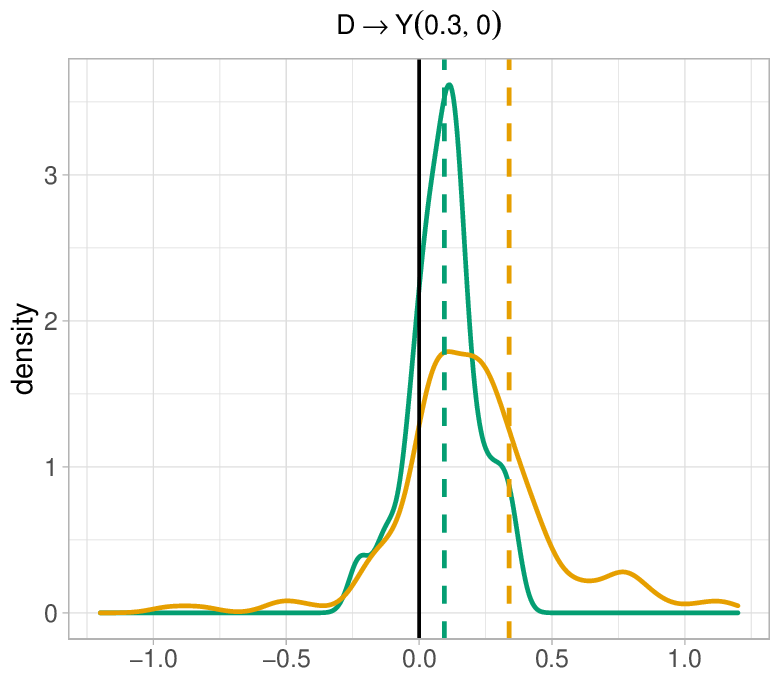}
\includegraphics[width=0.48\textwidth,height=5.2cm]{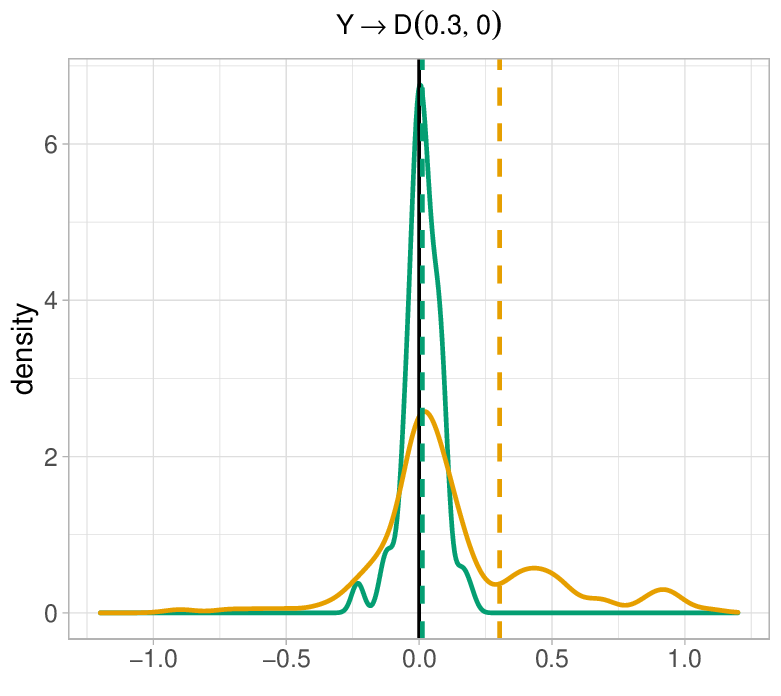}
\caption{
	IVW estimates based on a dataset simulated from a GWAS of BMI ($D$) and  a GWAS of fasting glucose ($Y$) with $(\betady,\betayd)=(0,0.3)$ (top) and $(\betady,\betayd)=(0.3,0)$ (bottom). Left panels: density plots for the normalized $\hat{\omega}_{Y,j}\hat{\gam}_{Y,j}/\hat{\gam}_{D,j}, j \in \H_{\dy}^{(0)}$ with $\tau_f=1.5$ (green,  for the focused IVW) or $\tau_f=\infty$ (orange,  for the overall IVW); dashed lines are the corresponding values of $\hat{\psi}_{\dy}$. Right panels: density plots for the normalized $\hat{\omega}_{D,j}\hat{\gam}_{D,j}/\hat{\gam}_{Y,j}, j \in \H_{\yd}^{(0)}$ with $\tau_f=1.5$ (green, for the focused IVW) or $\tau_f=\infty$ (orange,   for the overall IVW); dashed lines are the corresponding values of $\hat{\psi}_{\yd}$. 
}
\label{fig-ivw}
\end{figure}

In Figure \ref{fig-ivw}, we illustrate the effect of focusing based on  a dataset simulated from a  GWAS for body mass index ($D$) and a GWAS for fasting glucose ($Y$). The proportions of SNPs in four subsets are $(\rho_{null},\rho_{\dy},\rho_{\yd},\rho_{pl})=(0.15, 0.29, 0.29, 0.27)$, respectively, where $\rho_{*}=|\V_{*}|/p$ for $*\in \{null, \dy, \yd, pl\}$. More  details of configurations and implementations are presented in Section \ref{sec-simu-1}. Let $\widehat{\mathcal{S}}_D=\{j: |\hat{\gam}_{D,j}|\geq \tau_s\xi_{D,j}\}$ be the estimated set of relevant SNPs.
Under the null $\betady=0$ (top left), only 44\% of SNPs in $\widehat{\mathcal{S}}_D$ have $\pi_{Y,j}=0$, and the overall IVW estimator (the IVW estimator based on all SNPs in $\widehat{\mathcal{S}}_D$) is far away from the truth $\betady=0$ and can lead to false rejection of the null. In comparison, 79\% of the SNPs in $\widehat{\F}_{\dy}$ has $\pi_{Y,j}=0$, and the focused IVW estimator  (the IVW estimator based on SNPs in  $\widehat{\F}_{\dy}$) is close to the truth $\betady=0$ and does not reject the null. Under the alternative with $\betady\neq 0$  (top right), the focused IVW is closer to zero compared to the overall IVW, but both test statistics significantly deviate from the null. A similar story holds for the other direction based on the plots in the second row.

\subsection{Limiting distribution and rejection regions}
\label{sec3-3}
We establish the limiting distribution of the proposed  focused IVW estimator $\hat{\psi}_{\dy}$  in (\ref{psi-dy}). Based on that, we further derive the rejection region which controls Type I error at the nominal level. Method and theoretical results for the focused Median estimator are presented in the supplements.

\begin{condition}[Noise distribution of the effect estimates]
\label{cond1}
Assume that for each $j=1,\dots,p$, $\hat{\gam}_{D,j}-\gam_{D,j}=\frac{1}{n_D}\sum_{i=1}^{n_D}\delta_{i,j}^{(D)}+rem_{D,j}$ where $\delta_{i,j}^{(D)}$ are independent sub-Gaussian random variables with mean zero and the remainder terms satisfy that $\max_{j\leq p}|rem_{D,j}|=o_P(n_D^{-1/2})$. For each $j=1,\dots,p$, $\hat{\gam}_{Y,j}-\gam_{Y,j}=\frac{1}{n_Y}\sum_{i=1}^{n_Y}\delta_{i,j}^{(Y)}+rem_{Y,j}$ where $\delta_{i,j}^{(Y)}$ are independent sub-Gaussian random variables with mean zero and the remainder terms satisfy that $\max_j|rem_{Y,j}|=o_P(n_Y^{-1/2})$. Moreover, all the elements in $(\hat{\bgam}^T_{Y},\hat{\bgam}^T_{D})$ are mutually independent.  Both $\Vdy$ and $\Vyd$ are non-empty.
\end{condition}
%Condition \ref{cond1} assumes sub-Gaussian noises for two MR studies and the effects sizes estimates are independent of each other. Linkage disequilibrium pruning or clumping can produce a set of SNPs with mild correlations. 

 Define $z_{[a,b]}$ as a truncated standard normal random variable bounded by $[a,b]$. That is, $z_{[a,b]}$ has density $\phi(x)/[\Phi(a)-\Phi(b)]$ for $a<x<b$ and has density zero anywhere else, where $\phi(\cdot)$ and $\Phi(\cdot)$ are respectively the density function and cumulative distribution function of standard normal distribution. Theorem \ref{thm-type1} derives the limiting distribution of $\hat{\psi}_{\dy}$ under the null.  
\begin{theorem}[Limiting distribution under the null]
\label{thm-type1}
Suppose  that 
$\betady\betayd\neq 1$, equation \eqref{eq-gam}, and Condition \ref{cond1} hold. Let $\tau_f\in[c_0,\sqrt{2\log p}]$ for some constant $c_0>0$. Assume that
\begin{align}
\label{strong-invalid}
\min_{j\in\Vyd\cup\V_{pl}} |\pi_{Y,j}/\xi_{Y,j} | \geq c_1\tau_f\sqrt{\log p}
\end{align}
for some large enough constant $c_1$.
Then the focused set satisfies $\P_{\H_{\dy}^{(0)}}(\widehat{\F}_{\dy}\cap (\Vyd\cup \V_{pl})=\emptyset)\rightarrow 1$. \\
In the event that  $\{\log p\lesssim |\widehat{\F}_{\dy}|\ll n_Y/\tau_f^2,~\max_{j\in\widehat{\F}_{\dy}}\hat{\omega}_{Y,j} =o(\sum_{j\in\widehat{\F}_{\dy}}\hat{\omega}_{Y,j})\}$,  we have
\begin{align}
\P_{\H_{\dy}^{(0)}}\left(\hat{\psi}_{\dy}\leq t |\widehat{\F}_{\dy},\hat{\bgam}_{D}\right)&=\Phi\left(\frac{t}{\sqrt{var(z_{[-\tau_f,\tau_f]})/(\sum_{j\in\widehat{\F}_{\dy}}\hat{\omega}_{Y,j})}}\right)+ o(1),\label{dist-null}
\end{align}
where $var(z_{[-\tau_f,\tau_f]})=1-2\tau_f\phi(\tau_f)/\{\Phi(\tau_f)-\Phi(-\tau_f)\}$ as $(n_Y, p)\rightarrow\infty$. 
\end{theorem}
The first part of Theorem \ref{thm-type1} states that the focused set $\widehat{\F}_{\dy}$ contains only the  non-pleiotropic IVs, $\V_{null}\cup\Vdy$,  with high probability. This result critically relies on condition (\ref{strong-invalid}),   which says that the nonzero pleiotropic effects are relatively large in comparison to the noise level. It is  needed to distinguish valid IVs from invalid ones; see, for example, \citet{TSHT} and \citet{Guo:2018aa} who also makes this assumption. Some genetics literature has studied the magnitude of effect sizes. For instance, \citet{zhang2018estimation} finds that SNPs tend to have larger effect sizes on the early growth traits and many common disease traits. Hence, the direct effects of SNPs on these traits are likely to be larger. 

In the second part of Theorem \ref{thm-type1}, we prove that the limiting distribution of $\hat{\psi}_{\dy}$ is asymptotically normal with mean zero and variance $var(z_{[-\tau_f,\tau_f]})/(\sum_{j\in\widehat{\F}_{\dy}}\hat{\omega}_{Y,j})$ under the null. The conditions on $|\widehat{\F}_{\dy}|$ and $ \{\widehat{\omega}_{Y,j}\}_{j\in\widehat{\F}_{\dy} } $ are used to verify Lyapunov's condition for central limit theory. They are relatively mild and hold if $\log p\lesssim|\widehat{\F}_{\dy}|\ll n_Y/\log p$ and the distribution of IV strengths is not too uneven.
We note that Theorem \ref{thm-type1} does not require the IVs being strongly associated with the exposure $D$ and thus having weak IVs does not affect the type I error. We will show later that IV strengths do affect the power.

Theorem \ref{thm-type1} motivates the following rejection region when testing $\H_{\dy}^{(0)}$ based on $\hat{\psi}_{\dy}$,
\begin{align}
\label{rr}
 \mathcal{R}_{\dy}(\alpha):=\left\{\hat{\psi}_{\dy}: |\hat{\psi}_{\dy}| \geq q_{1-\alpha/2}\sqrt{\frac{var(z_{[-\tau_f,\tau_f]})}{\sum_{j\in\widehat{\F}_{\dy}}\hat{\omega}_{Y,j}}}\right\},
\end{align}
where $q_{\alpha}$ is the $\alpha$-th quantile of the standard normal distribution.
We summarize the proposed testing procedure in Algorithm \ref{alg-test}. The Type I error guarantee is formally stated in Corollary \ref{cor-err}.

\begin{algorithm}[H]
	\SetKwInOut{Input}{Input}
	\SetKwInOut{Output}{Output}
	\Input{$(\hat{\bm \gam}_D,\hat{\bm \gam}_Y)$, their standard errors $\xi_{D,j}$ and $\xi_{Y,j}$,  $j=1,\dots, p$, significance level $\alpha$, a screening threshold level $\tau_s$ and a focused threshold level $\tau_f\in[c_0,c_0\sqrt{2\log p}]$.}
	
	1. Compute $\widehat{\F}_{\dy}$ via (\ref{eq-Fdy}).
	
	2. If $\widehat{\F}_{\dy}=\emptyset$, then reject $\mathcal{H}_{\dy}$. Otherwise, compute $\hat{\psi}_{\dy}$ as in (\ref{psi-dy}).
	
	%\For{ $b=1,\dots, B$}{
		%For $j\in\widehat{\F}_{\dy}$, independently generate $t^*_{j,b}\sim z_{[-\tau_f,\tau_f]}$ and compute
		%\[
		%  \hat{\psi}_{\dy,b}^*=\frac{\sum_{j\in\widehat{\F}_{\dy}}\hat{\omega}_{Y,j}\xi_{Yj}t^*_{j,b}/\hat{\gam}_{D,j}}{\sum_{j\in\widehat{\F}_{\dy}}\hat{\omega}_{Y,j}}.
		%\]
		%}
	%Let $\widehat{G}_{1:B}^*(u)$ denote the empirical distribution of $\hat{\psi}_{\dy,b}^*$, $b=1,\dots, B$. Compute the rejection region via
	%\begin{align}
	%\label{rr-est1}
	%  \widehat{\mathcal{R}}_{\dy}=\left\{u\in\R: \widehat{G}_{1:B}^*(u)\leq \alpha/2~\text{or}~\widehat{G}_{1:B}^*(u)\geq 1-\alpha/2\right\}.
	%\end{align}
	
	3. Reject $\mathcal{H}_{\dy}$ if $\hat{\psi}_{\dy}\in \widehat{\mathcal{R}}_{\dy}(\alpha)$ for $\widehat{\mathcal{R}}_{\dy}(\alpha)$ defined in (\ref{rr}). 	
	\caption{Testing $\mathcal{H}_{\dy}$ based on the focused IVW at significance level $\alpha$.}
	\label{alg-test}
\end{algorithm}

\begin{corollary}[Type I error guarantee]
\label{cor-err}
Under the conditions in Theorem \ref{thm-type1}, the rejection region defined in (\ref{rr}) satisfies $\P_{\H^{(0)}_{\dy}}\big(\hat{\psi}_{\dy}\in\mathcal{R}_{\dy}(\alpha)\big)=\alpha+o(1)$ for any $0<\alpha<1$.
\end{corollary}

Finally, we remark that the results in Theorem \ref{thm-type1} and Corollary \ref{cor-err} account for the effect of post-selection and the limiting distribution in (\ref{dist-null}) is based on the truncated normal density rather than the standard normal density. Alternatively, if $\widehat{\F}_{\dy}$ is computed based on samples independent of $(\hat{\bm\gam}_D, \hat{\bm\gam}_Y)$, 
then a simpler limiting distribution can be derived. In practice, we may not have proper external datasets to select the focused sets for both directions. Hence, we perform selection and testing in the same data set and adjust for the post-selection effect. We derive  results parallel to Corollary \ref{cor-err} with external selection datasets in the supplements.

\subsection{Power analysis}
\label{sec3-3}
We derive the power of the proposed test and prove that the test is asymptotically unbiased under mild conditions. Define  the signal-to-noise ratio (SNR) for $\hat{\gam}_{Y,j}$ as $\mu_j=\gam_{Y,j}/\xi_{Y,j}$ and let 
\[
  \Gamma_Y^+=\{j\leq p: |\mu_j|\geq \tau_f\sqrt{\log p}\}~\text{and}~\Gam_Y^-=\{j\leq p: |\mu_j|\leq c_0\tau_f\}
\]
for some small enough constant $0<c_0<0.5$. Let $z_{[a,b]}(\mu)$ be a truncated normal with mean $\mu$ and unit variance in the interval $[a,b]$, i.e., its density is $\phi(x-\mu)/[\Phi(b)-\Phi(a)]$. Theorem \ref{lem-power} derives the limiting distribution of the focused IVW estimator $\hat{\psi}_{\dy}$ for any $\betady$.

\begin{theorem}[Limiting distribution under the alternative]
\label{lem-power} Suppose  that 
	$\betady\betayd\neq 1$, equation \eqref{eq-gam}, and Condition \ref{cond1} hold. 
Assume $\Gam_Y^+\cup\Gam_Y^-=\{1,\dots,p\}$. In the event that $\{\log p\lesssim|\widehat{\F}_{\dy}|\ll n_Y/\tau_f^2,~\max_{j\in\widehat{\F}_{\dy}}\hat{\omega}_{Y,j} =o(\sum_{j\in\widehat{\F}_{\dy}}\hat{\omega}_{Y,j})\}$, we have 
\begin{align*}
    \P(\hat{\psi}_{\dy}\leq t|\widehat{\F}_{\dy},\hat{\bgam}_D)=\Phi\left(\frac{t-\mu_{\dy}}{\sig_{\dy}}\right)+o(1),
\end{align*}
for any $t\in\R$ as $(n_Y,p)\rightarrow\infty$, where
\begin{align*}
   & \mu_{\dy}=\frac{\sum_{j\in\widehat{\F}_{\dy}}  \hat{\omega}_{Y,j}\hat{\gam}^{-1}_{D,j}\xi_{Y,j}\E[z_{[-\tau_f,\tau_f]}(\mu_j)]}{\sum_{j\in\widehat{\F}_{\dy}}\hat{\omega}_{Y,j}}~\text{and}~ \sig^2_{\dy}=\frac{\sum_{j\in\widehat{\F}_{\dy}}\hat{\omega}_{Y,j}var(z_{[-\tau_f,\tau_f]}(\mu_j))}{(\sum_{j\in\widehat{\F}_{\dy}}\hat{\omega}_{Y,j})^2}.
\end{align*}
\end{theorem}
In Theorem \ref{lem-power}, we show that $\hat{\psi}_{\dy}$ is asymptotically normal under the alternative. The condition $\Gam_Y^+\cup\Gam_Y^-=\{1,\dots,p\}$ means that the SNR for $\hat{\bgam}_Y$ can be divided to two sets $  \Gamma_Y^+$ and $  \Gamma_Y^-$, and there is no $|\mu_j|$ close to the threshold $\tau_f$. Theorem \ref{lem-power} considers the case where $|\widehat{\F}_{\dy}|\rightarrow\infty$. If $\widehat{\F}_{\dy}=\emptyset$, we always reject the null and the power is one in this case. 

 As a consequence of post-selection, the asymptotic mean and asymptotic variance are both functions of $\mu_j,~j\in\widehat{\F}_{\dy}$. The limiting distribution under the null in Theorem \ref{thm-type1} is a special case with $\mu_j=0,~j\in\widehat{\F}_{\dy}$ in the above expression. Under the alternative, $\mu_j$'s are not all zero in $\widehat{\F}_{\dy}$. As a result, $\mu_{\dy}$ is generally nonzero but $\sig_{\dy}^2$ can be smaller than the variance under the null. The dependence of $\sig_{\dy}$ on $\mu_j$'s complicates the power as a function of $\mu_{\dy}$. 

In the next theorem, we consider a simplified scenario with approximately equal $\mu_j$ for $j\in\Gam_Y^-$.
\begin{lemma}[A simple scenario with approximately equal SNR]
\label{thm2-power}
Assume the conditions of Theorem \ref{lem-power} and $\tau_f\geq c_0>0$ for some large enough constant $c_0$, and 
\begin{align}
\label{cond-power}
  \max_{j\in\Gam_Y^-}|\mu_j-\bar{\mu}|=o(|\Gam_Y^-|^{-1/2})~\text{for}~\bar{\mu}=\frac{\sum_{j\in\Gam_Y^-}\mu_j}{|\Gam_Y^-|}.
\end{align}
In the event that $\big\{\max_{j\in\widehat{\F}_{\dy}}\hat{\omega}^{1/2}_{Y,j}\lesssim\sum_{j\in\widehat{\F}_{\dy}}\hat{\omega}_{Y,j}^{1/2}/|\widehat{\F}_{\dy}|, |\bar{\mu}|\gg \sqrt{\log(1/\alpha)/|\widehat{\F}_{\dy}|}\big\}$ for some constant $C>0$, the power function $\P(\hat{\psi}_{\dy}\in\widehat{\mathcal{R}}_{\dy})$ is an increasing function of $|\bar{\mu}|$ as $(p,n_Y)\rightarrow \infty$. 
As a consequence, the proposed test is asymptotically unbiased, i.e., $\P_{\betady}(\hat{\psi}_{\dy}\in\widehat{\mathcal{R}}_{\dy})\geq \alpha-o(1)$ for any $\betady\neq 0$.
\end{lemma}
Lemma \ref{thm2-power} shows that under certain conditions, the proposed test is asymptotically unbiased. The condition (\ref{cond-power}) is needed to simplify the technicality in the proof. In the simulation and real data, our proposal is always empirically unbiased and has desirable power in most cases.
The results of Theorem \ref{thm-type1} and Lemma \ref{thm2-power} allow for any small $\alpha$. Hence, leveraging the proposed method in multiple testing has the same theoretical guarantees. 
%Second, it may not be very realistic as the null IVs, $j\in\V_{null}$ has $\mu_j=0$ and thus is a subset of $\Gam_Y^-$. Indeed, if we take $\tau_s\geq c_1\sqrt{\log p}$ with a large enough constant $c_1$, then $j\in \V_{null}$ cannot be selected into $\widehat{\F}_{\dy}$ with high probability. Then condition (\ref{cond-power}) can be changed in a way that $\Gam_Y^-$ can be replaced with $\Gam_Y^-\setminus \V_{null}$, which may be more plausible.

\section{Numerical Experiments}
\label{sec-simu}
We perform multiple numerical studies with data simulated from four GWASs. We evaluate the Type I error and power performance of the focused IVW and focused Median estimators proposed in Section \ref{sec3}, and two other benchmark methods that can deal with invalid IVs, the MR-Median \citep{Bowden16} and MR-Egger \citep{Bowden15, Kole15}. To evaluate the effect of different thresholds, we report the results for the focused IVW and focused Median with $\tau_f=1.2$ and $\tau_f=1.5$. We set the relevant SNP threshold as $\tau_s=\Phi^{-1}(1-1/p)$ for all the methods where $p$ is the total number of candidate IVs. Following the common practice, for each direction, the MR-Median and MR-Egger methods are performed using SNPs that pass the threshold $\tau_s$. R code for implementing all the methods is available at \url{https://github.com/saili0103/BidirectionalMR}.

\subsection{ Simulated data from GWAS of BMI and fasting glucose}
\label{sec-simu-1}

To simulate data that closely mirror the real GWASs, we first obtain a GWAS for BMI and a GWAS for fasting glucose from the IEU OpenGWAS project at  (\url{https://gwas.mrcieu.ac.uk}) with GWAS ID \texttt{ieu-b-40} and \texttt{ebi-b-114} \citep{elsworth2020mrc}. The sample sizes for both GWASs are around $10^5$. After  linkage disequilibrium (LD) pruning, we obtain $p=394$ SNPs with LD$<10^{-5}$ and their associations with at least one trait has $p$-value $<10^{-3}$.  Let $\widehat{\bm \alpha}^{(\textup{bmi})}$ denote the effect size vector from the BMI GWAS and $\widehat{\bm \alpha}^{(\textup{fg})}$ denote the effect size vector from the GWAS on fasting glucose. Then we generate $\pi_{Y,j}=\widehat{\alpha}^{(\textup{fg})}_j\textup{Ber}_j(p_{Y,j})$, with $p_{Y,j}=\textup{rank}(|\widehat{\alpha}^{(\textup{fg})}_j|/\xi_{Y,j})/p$ to reflect that $\pi_{Y,j}$ with a small $|\hat{\pi}_{Y,j}|/\xi_{Y,j}$ are more likely to be zero, where $\textup{Ber}_j(p_{j})$ is an independent Bernoulli random variables with probability $p_j$.  We generate $\pi_{D,j}$'s similarly based on $\widehat{\bm \alpha}^{(\textup{bmi})}$. This procedure gives IV proportions $(\rho_{null},\rho_{\dy},\rho_{\yd},\rho_{pl})\approx(0.15, 0.29, 0.29, 0.27)$. The correlation between $\bpi_D$ and $\bpi_Y$ is 0.17.

\begin{figure}[H]
\includegraphics[width=0.99\textwidth,height=4.5cm]{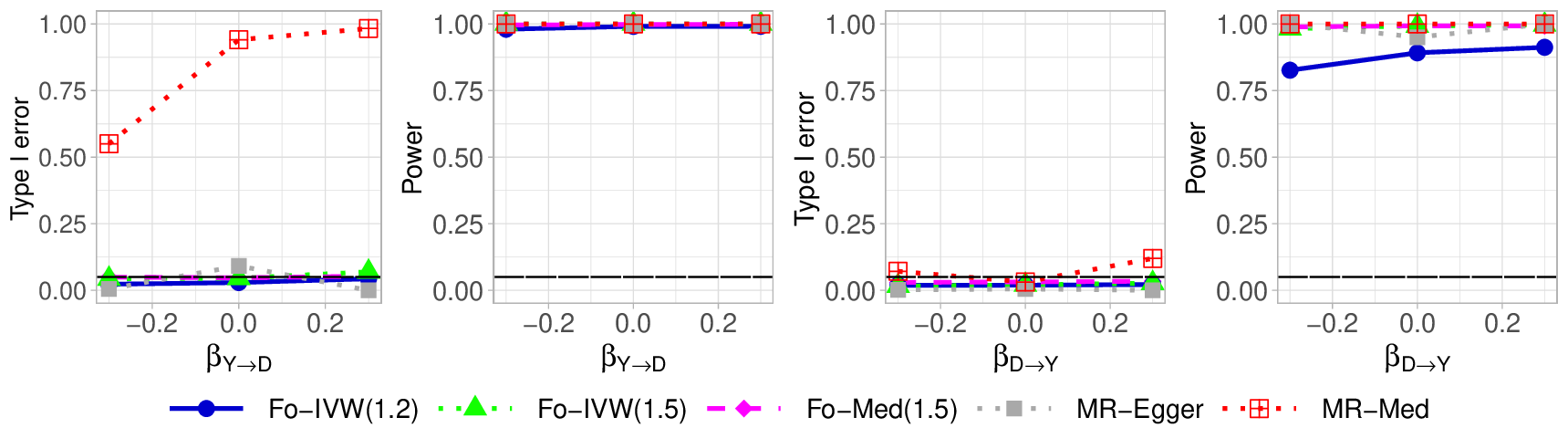}
\includegraphics[width=0.99\textwidth,height=4.5cm]{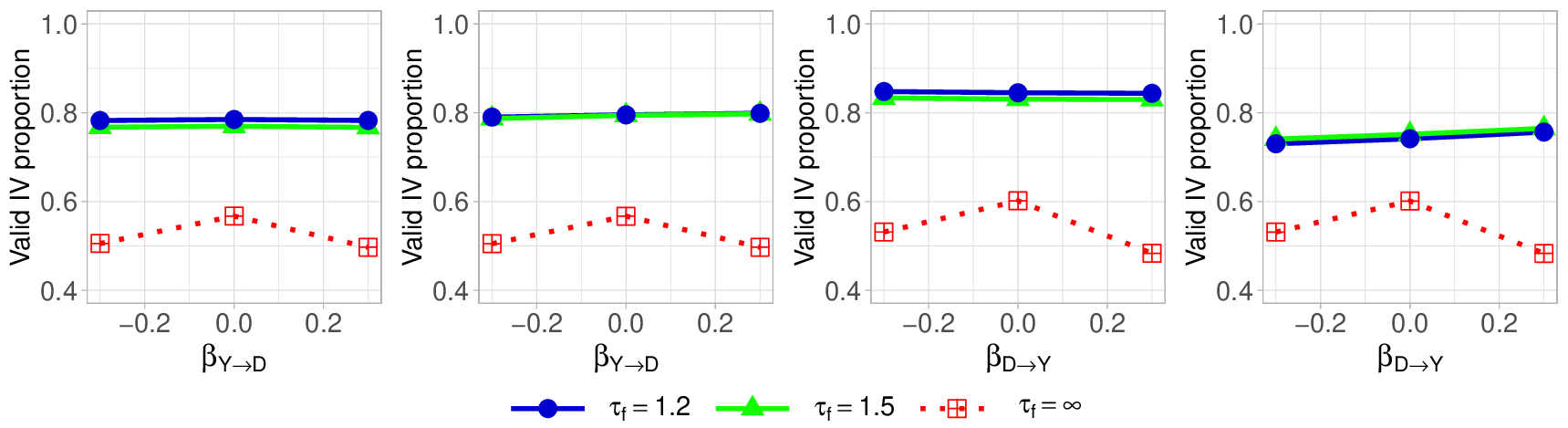}
\caption{First row: rejection rate of $\H_{\dy}^{(0)}$ at significance level $\alpha=0.05$ with $\betady=0$ (left) and $\betady=0.3$ (left middle) and rejection rate of $\H_{\yd}^{(0)}$ with $\betayd=0$ (right middle) and $\betayd=0.3$ (right). Second row: the proportion of valid IVs in the focused sets with $\tau_f=1.2, \tau_f=1.5,$ and  $\tau_f = \infty$ (for the overall methods, i.e., MR-Median and MR-Egger) in each scenario. Results are from 3000 independent experiments generated from a GWAS for BMI and a GWAS for fasting glucose with $(\rho_{null},\rho_{\dy},\rho_{\yd},\rho_{pl})\approx(0.15, 0.29, 0.29, 0.27)$.}
\label{fig1-1}
\end{figure}

The results are in Figure \ref{fig1-1}. We see that the focused IVW, focused Median, and MR-Egger have empirical Type I error rates close to the nominal level but the conventional MR-Median has severe Type I error inflation for testing $\H_{\dy}^{(0)}=0$. 
	Type I error inflation using MR-Median has also been noted in \citet{Bowden16} when there exists directional pleiotropy or there are more than 30\% of invalid IVs.
When the alternative hypotheses are true, all the methods have power close to one except for the focused IVW with $\tau_f=1.2$. 
This agrees with our previous discussion that larger $\tau_f$ can have better power. From the second row, we see the the proportion of valid IVs, i.e., those in $\Vdy$ for testing $\H_{\dy}^{(0)}$ are significantly improved with focusing. Higher valid IV proportions leads to smaller bias induced by invalid IVs and better control of Type I error.

\begin{figure}[H]
\includegraphics[height=4.5cm,width=0.99\textwidth]{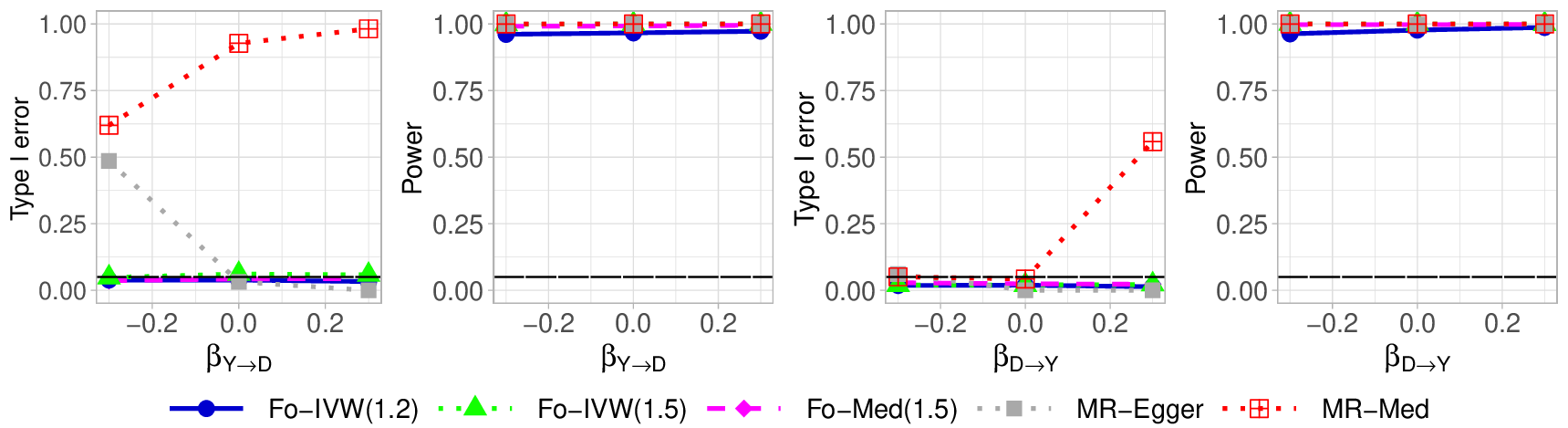}
\includegraphics[height=4.5cm,width=0.99\textwidth]{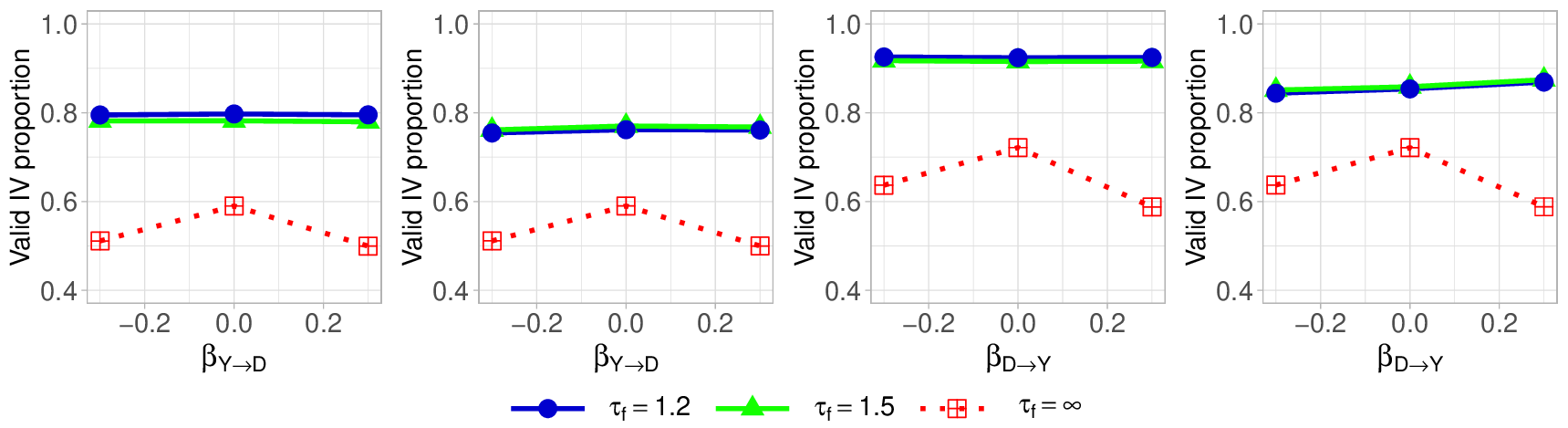}
\caption{First row: rejection rate of $\H_{\dy}^{(0)}$ at significance level $\alpha=0.05$ with $\betady=0$ (left) and $\betady=0.3$ (left middle) and rejection rate of $\H_{\yd}^{(0)}$ with $\betayd=0$ (right middle) and $\betayd=0.3$ (right). Second row: the proportion of valid IVs in the focused sets with $\tau_f=1.2, \tau_f=1.5,$ and  $\tau_f = \infty$ (for the overall methods, i.e., MR-Median and MR-Egger) in each scenario. Results are from 3000 independent experiments generated from a GWAS for BMI and a GWAS for fasting glucose with $(\rho_0,\rho_{\dy},\rho_{\yd},\rho_{pl})\approx(0.17,0.25, 0.37,0.21)$.}
\label{fig1-2}
\end{figure}

In the next experiment, we evaluate these methods with different IV proportions. In the results reported in Figure \ref{fig1-2}, we generate $\pi_{Y,j}=\widehat{\alpha}^{(\textup{fg})}_j\textup{Ber}_j(p_{Y,j}^{0.7})$ for $p_{Y,j}=\textup{rank}(|\widehat{\alpha}^{(\textup{fg})}_j|/\xi_{Y,j})/p$ defined as above.
Let $\pi_{D,j}$ are generated similarly such that $\pi_{D,j}=\widehat{\alpha}^{(\textup{bmi})}_j\textup{Ber}_j(p_{D,j}^{0.7})$. It gives IV proportions $(\rho_0,\rho_{\dy},\rho_{\yd},\rho_{pl})\approx(0.17,0.25, 0.37,0.21)$. The correlation between $\bpi_D$ and $\bpi_Y$ is 0.13.  This setting is more challenging for testing $\H_{\dy}^{(0)}$ than for testing $\H_{\yd}^{(0)}$ as $\rho_{\dy}$ is smaller. We see that both the Egger and overall Median have severely inflated Type I errors for testing $\H_{\dy}^{(0)}$.

\subsection{ Simulated data from GWAS of LDL and CAD}
\label{sec-simu-2}
In this section, we simulate data from a different pair of real data, a GWAS of low-density lipoprotein (LDL) and a GWAS of coronary artery disease (CAD), with GWAS ID \texttt{ieu-b-110} and \texttt{ebi-a-GCST005195}, respectively.  There are $p=1332$ candidate IVs in total after LD clumping and $p$-value screening. We denote LDL by $D$ and CAD by $Y$ and the true parameters are generate in the same way as in Section \ref{sec-simu-1}.

\begin{figure}[H]
	\includegraphics[width=0.99\textwidth,height=4.5cm]{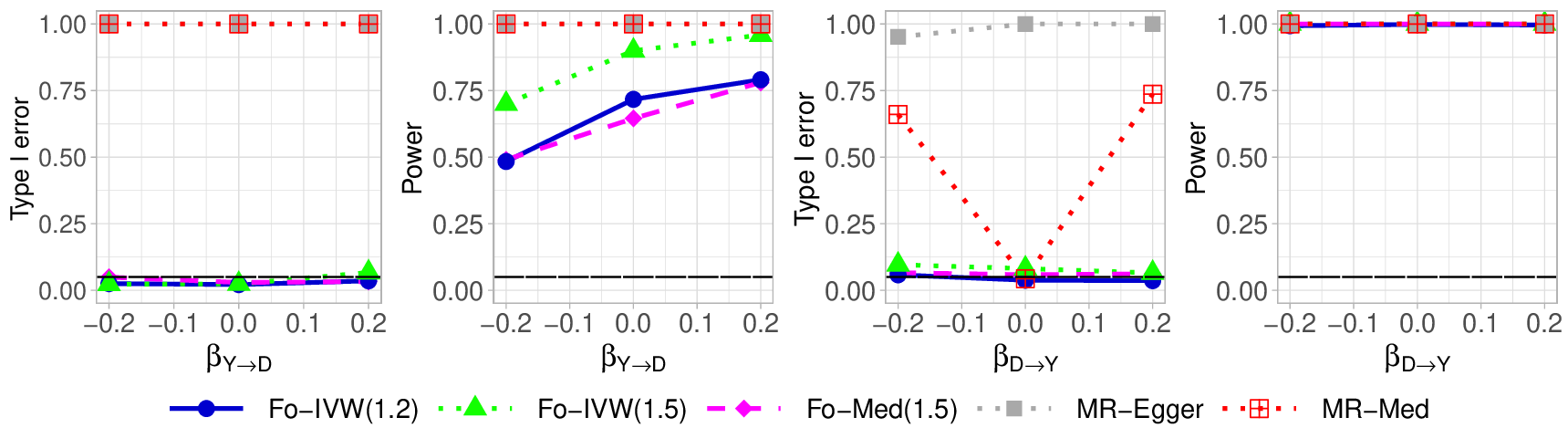}
	\includegraphics[width=0.99\textwidth,height=4.5cm]{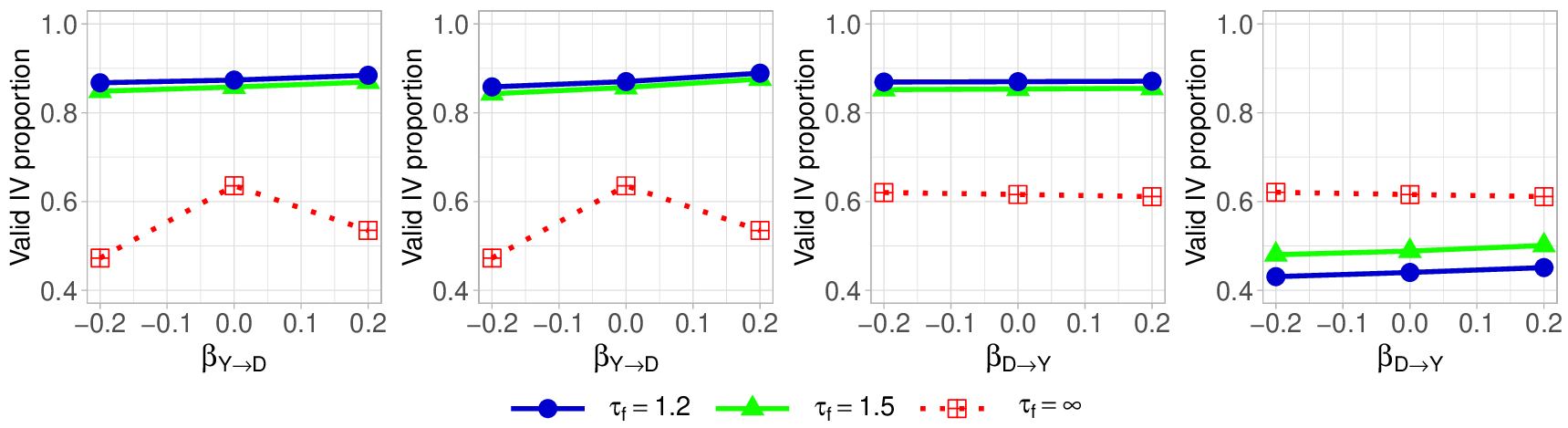}
	\caption{First row: rejection rate of $\H_{\dy}^{(0)}$ at significance level 0.05 with $\betady=0$ (left) and $\betady=0.2$ (left middle) and rejection rate of $\H_{\yd}^{(0)}$ with $\betayd=0$ (right middle) and $\betayd=0.2$ (right). Second row: the proportion for valid IVs in the focused sets ($\tau_f=\infty$ for overall Median and Egger) in each corresponding scenario.  3000 experiments are simulated base on GWAS for LDL and CAD with $(\rho_{null},\rho_{\dy},\rho_{\yd},\rho_{pl})\approx(0.20,0.30,0.28,0.22)$.}
	\label{fig2-1}
\end{figure}

In Figure \ref{fig2-1}, we consider a setting where $(\rho_{null},\rho_{\dy},\rho_{\yd},\rho_{pl})\approx(0.20,0.30,0.28,0.22)$.
From Figure \ref{fig2-1}, we see that both the conventional MR-Median and  MR-Egger fail to control Type I errors. The failure of MR-Egger is mainly because the generated $\bpi_D$ and $\bpi_Y$ have large correlation (about $ 0.23$) in the current setting. The focused methods have Type I error close to the nominal level and we see the focused Median is more robust than focused IVW when $\tau_f=1.5$ for testing $\H_{\yd}^{(0)}$. In terms of power, the focused Median is less powerful than focused IVW under the same $\tau_f$. This aligns with the efficiency comparison between IVW and MR-Median estimators in the existing literature. In terms of the valid IV proportions, we see that they are improved under the null by focusing. Under the alternative, the valid IV proportions can be even lower with focusing. We mention that this observation does not conflict with our analysis as our theory only justifies removing the  invalid IV under the null. Under the alternative, as we see in Figure \ref{fig2-1}, the power of the focused methods can still be close to one with a small proportion of valid IVs.

In the supplements, we report the results of another setting based on the GWAS for LDL and CAD with $(\rho_{null},\rho_{\dy},\rho_{\yd},\rho_{pl})\approx(0.16, 0.24, 0.33, 0.27)$. Similar phenomena are observed in that setting.

\section{Real Data Examples}
\label{sec-data}

We apply our methods to test causal relationships using public GWASs. We consider five commonly studied phenotypes: body mass index (\texttt{bmi}), systolic blood pressure (\texttt{sbp}), type-2 diabetes (\texttt{t2d}), coronary artery disease (\texttt{cad}), and education (\texttt{edu}). The summary statistics are obtained from ``IEU OpenGWAS project'' \citep{elsworth2020mrc}. The preprocessing steps and GWAS IDs are in the supplements.

\begin{figure}[H]
\includegraphics[width=0.495\textwidth, height=6cm]{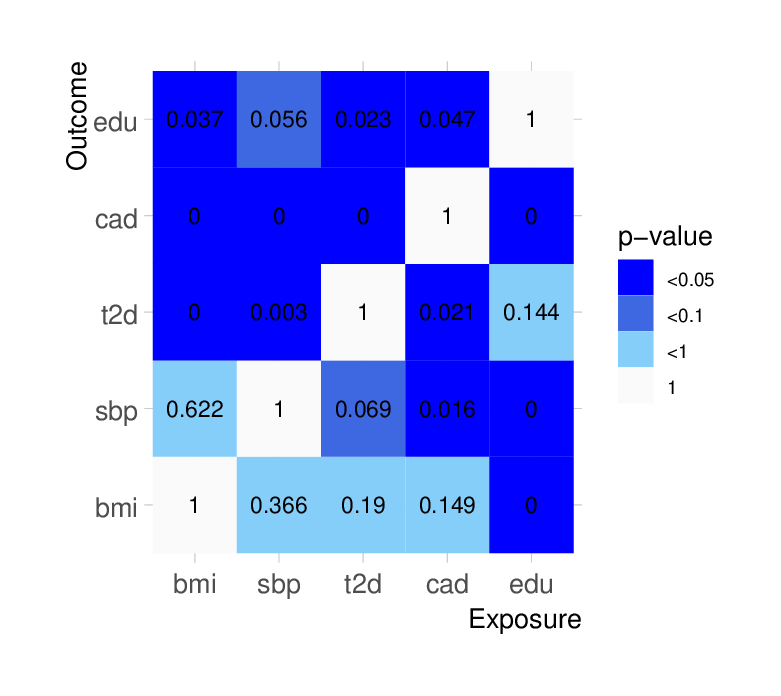}
\includegraphics[width=0.495\textwidth, height=6cm]{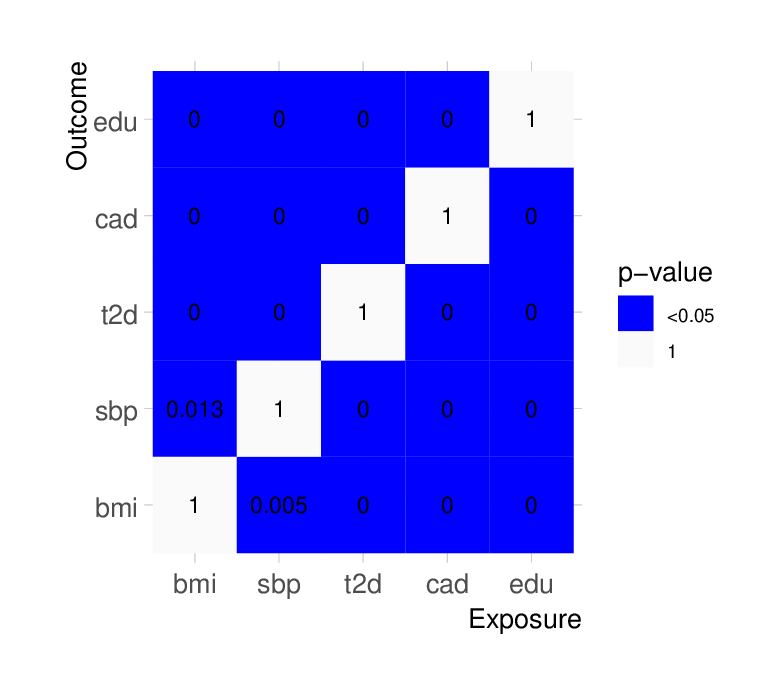}
\includegraphics[width=0.495\textwidth,height=6cm]{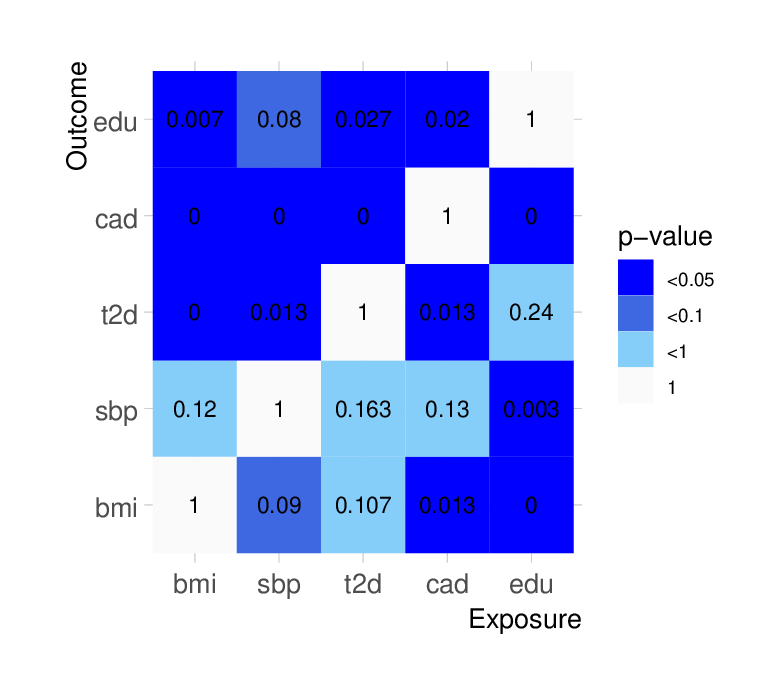}
\includegraphics[width=0.495\textwidth,height=6cm]{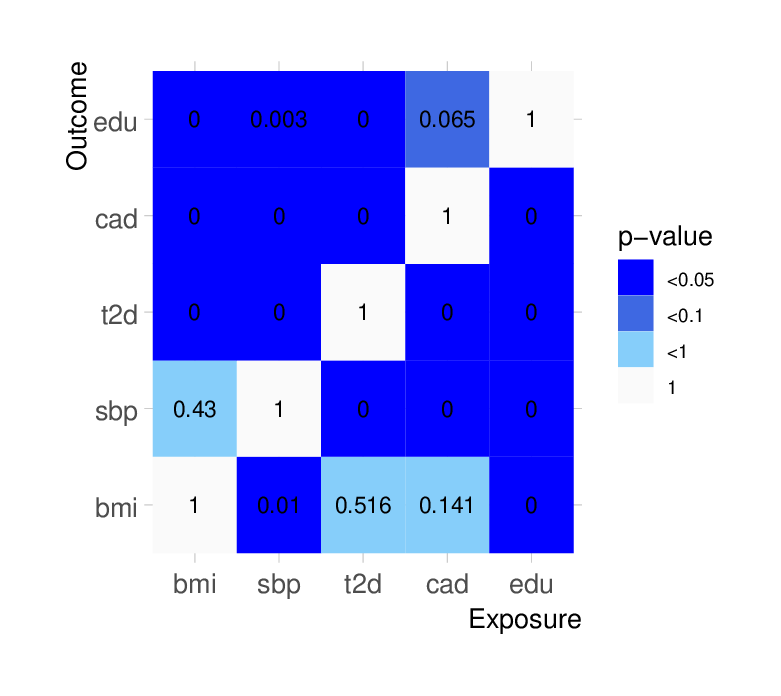}
\caption{Bi-directional hypothesis testing based on focused IVW with $\tau_f=1.5$ (top left), overall IVW (top right), focused Median with $\tau=1.5$ (bottom left)  and the overall Median (bottom right). The $p$-values are from testing $\H_{\dy}^{(0)}$ where $x$-axis corresponds to the exposure and $y$-axis corresponds to the outcome.}
\label{fig-data}
\end{figure}

For each pair of these five traits, we test the bi-directional causal effects and report the $p$-values in Figure \ref{fig-data}. We compare the focused IVW ($\tau_f=1.5$) with the overall IVW and compare the focused Median ($\tau_f=1.5$) with MR-Median. We set $\tau_s=\Phi^{-1}(1-1/p)$ for all the methods when selecting relevant SNPs. Results based on the MR-Egger are given in the supplements. 
We highlight the following observations.

First, the focused IVW and focused Median successfully detects the causal effects of \texttt{bmi} on \texttt{t2d}, \texttt{bmi} on \texttt{cad}, \texttt{sbp} on \texttt{t2d}, and \texttt{sbp} on \texttt{cad} at 5\% significance level. These four effects are considered causal in the literature \citep{morrison2020ng}.  The reverse of these four relationships have been considered as implausible in the existing literature. 
We see that the focused Median and focused IVW both have three non-significant $p$-values for these four reverse tests at 5\% significance level. These results show the robustness of the focused methods in detecting causal directions.

Second, we find significant bi-directional causal effects between \texttt{bmi} and \texttt{edu} at 5\% significance level. \citet{heckman2018returns} and many related works present significant causal effect of education on health, which agrees with the current analysis. For the reverse direction, childhood obesity has been found to be associated with lower academic performance \citep{devaux2019relationship} and psychological issues \citep{nieman2012psychosocial}, which may explain the causal effect of \texttt{bmi} on \texttt{edu}.

In comparison, the overall IVW and MR-Median report many more significant causal relationships. The results of IVW are dubious due to the prevalence of pleiotropy. The MR-Median reports significant causal effect of \texttt{t2d} on \texttt{sbp} and significant causal effect of \texttt{cad} on \texttt{sbp}.  The causal effects of diseases on the physical measures are barely supported by the existing literature and can be doubtful.
 The results given by Egger's method also has suspicious discoveries such as  \texttt{t2d}$\rightarrow$\texttt{sbp} and \texttt{cad}$\rightarrow$\texttt{sbp}. It fails to detect some causal effects, say, the effect of \texttt{edu} on health-related outcomes, which are well-established causal relationships.

To conclude, the focused methods give more reliable results in these real data experiments. They detected some well-known causal relationships and made less spurious causal discoveries. This again shows that focusing is more effective in eliminating invalid IVs and the results are less biased.
As multiple tests are conducted simultaneously, it is likely to make some false discoveries when setting the significance level at the nominal level 5\% for  each test. Taking the multiplicity correction into account, the focused methods still have more reliable results than their counterparts.

\section{Discussion}
\label{sec-diss}
This paper studies bi-directional causal inference with GWAS summary data. We first show that assumptions for common MR methods are often impossible or too stringent in the presence of a bi-directional relationship. We then propose a new focusing framework for testing bi-directional causal effects between two traits with possibly pleiotropic genetic variants, which can be coupled with many state-of-art MR methods and deliver reliable hypothesis testing results.

The focusing idea in this work can be generalized beyond two-sample MR. In generic IV studies, it may not be available a set of independent IVs, but the covariance of IVs is estimable. The dependence of IVs and the dependence of the effect estimates for two traits require more careful technical treatments.
Beyond the testing problem considered in this work, estimation and confidence interval construction for bi-directional causal effects have been barely studied with statistical guarantees. When having more than two phenotypes available, the current framework can also be generalized to learn causal networks, say, among gene expressions, under proper error criteria for multiple testing. 

\bibliographystyle{apalike}
\bibliography{reference_IV}

\begin{thebibliography}{}

\bibitem[Adam, 2019]{Adam:2019aa}
Adam, D. (2019).
\newblock The gene-based hack that is revolutionizing epidemiology.
\newblock {\em Nature}, 576(7786):196--199.

\bibitem[Bowden et~al., 2015]{Bowden15}
Bowden, J., Davey~Smith, G., and Burgess, S. (2015).
\newblock Mendelian randomization with invalid instruments: effect estimation
  and bias detection through egger regression.
\newblock {\em International journal of epidemiology}, 44(2):512--525.

\bibitem[Bowden et~al., 2016]{Bowden16}
Bowden, J., Davey~Smith, G., Haycock, P.~C., and Burgess, S. (2016).
\newblock Consistent estimation in mendelian randomization with some invalid
  instruments using a weighted median estimator.
\newblock {\em Genetic epidemiology}, 40(4):304--314.

\bibitem[Burgess et~al., 2015]{Burgess:2015ab}
Burgess, S., Small, D.~S., and Thompson, S.~G. (2015).
\newblock A review of instrumental variable estimators for mendelian
  randomization.
\newblock {\em Statistical Methods in Medical Research}, 26(5):2333--2355.

\bibitem[Carrasquilla et~al., 2021]{carrasquilla2021mendelian}
Carrasquilla, G.~D., Garc{\'\i}a-Ure{\~n}a, M., Fall, T., S{\o}rensen, T.~I.,
  and Kilpel{\"a}inen, T.~O. (2021).
\newblock Mendelian randomization suggests a bidirectional, causal relationship
  between physical inactivity and obesity.
\newblock {\em bioRxiv}.

\bibitem[Carreras-Torres et~al., 2018]{carreras2018role}
Carreras-Torres, R., Johansson, M., Haycock, P.~C., Relton, C.~L., Smith,
  G.~D., Brennan, P., and Martin, R.~M. (2018).
\newblock Role of obesity in smoking behaviour: Mendelian randomisation study
  in uk biobank.
\newblock {\em bmj}, 361.

\bibitem[Darrous et~al., 2021]{darrous2020simultaneous}
Darrous, L., Mounier, N., and Kutalik, Z. (2021).
\newblock Simultaneous estimation of bi-directional causal effects and
  heritable confounding from gwas summary statistics.
\newblock {\em Nature communications}, 12(1):1--15.

\bibitem[Davey~Smith and Ebrahim, 2003]{Davey03}
Davey~Smith, G. and Ebrahim, S. (2003).
\newblock Mendelian randomization: can genetic epidemiology contribute to
  understanding environmental determinants of disease?
\newblock {\em International journal of epidemiology}, 32(1):1--22.

\bibitem[Davey~Smith and Hemani, 2014]{DaveySmith14}
Davey~Smith, G. and Hemani, G. (2014).
\newblock {Mendelian randomization: genetic anchors for causal inference in
  epidemiological studies}.
\newblock {\em Human Molecular Genetics}, 23(R1):R89--R98.

\bibitem[Devaux and Vuik, 2019]{devaux2019relationship}
Devaux, M. and Vuik, S. (2019).
\newblock The relationship between childhood obesity and educational outcomes.
\newblock {\em OECD Health Policy Studies}.

\bibitem[Elsworth et~al., 2020]{elsworth2020mrc}
Elsworth, B., Lyon, M., Alexander, T., Liu, Y., Matthews, P., Hallett, J.,
  Bates, P., Palmer, T., Haberland, V., Smith, G.~D., et~al. (2020).
\newblock The mrc ieu opengwas data infrastructure.
\newblock {\em BioRxiv}.

\bibitem[Guo, 2021]{guo2021post}
Guo, Z. (2021).
\newblock Post-selection problems for causal inference with invalid
  instruments: A solution using searching and sampling.
\newblock arXiv:2104.06911.

\bibitem[Guo et~al., 2018a]{TSHT}
Guo, Z., Kang, H., Cai, T.~T., and Small, D.~S. (2018a).
\newblock Confidence intervals for causal effects with invalid instruments by
  using two-stage hard thresholding with voting.
\newblock {\em Journal of the Royal Statistical Society: Series B (Statistical
  Methodology)}, 80(4):793--815.

\bibitem[Guo et~al., 2018b]{Guo:2018aa}
Guo, Z., Kang, H., Cai, T.~T., and Small, D.~S. (2018b).
\newblock Confidence intervals for causal effects with invalid instruments by
  using two-stage hard thresholding with voting.
\newblock {\em Journal of the Royal Statistical Society: Series B (Statistical
  Methodology)}, 80(4):793--815.

\bibitem[Han, 2008]{Han:2008aa}
Han, C. (2008).
\newblock Detecting invalid instruments using {L1-GMM}.
\newblock {\em Economics Letters}, 101(3):285--287.

\bibitem[Hartwig et~al., 2017]{Bowden17}
Hartwig, F.~P., Davey~Smith, G., and Bowden, J. (2017).
\newblock Robust inference in summary data mendelian randomization via the zero
  modal pleiotropy assumption.
\newblock {\em International journal of epidemiology}, 46(6):1985--1998.

\bibitem[Hausman, 1983]{Hausman1983specification}
Hausman, J.~A. (1983).
\newblock Specification and estimation of simultaneous equation models.
\newblock {\em Handbook of econometrics}, 1:391--448.

\bibitem[Heckman et~al., 2018]{heckman2018returns}
Heckman, J.~J., Humphries, J.~E., and Veramendi, G. (2018).
\newblock Returns to education: The causal effects of education on earnings,
  health, and smoking.
\newblock {\em Journal of Political Economy}, 126(S1):S197--S246.

\bibitem[Hemani et~al., 2017]{hemani2017orienting}
Hemani, G., Tilling, K., and Davey~Smith, G. (2017).
\newblock Orienting the causal relationship between imprecisely measured traits
  using gwas summary data.
\newblock {\em PLoS genetics}, 13(11):e1007081.

\bibitem[Hemani et~al., 2018]{Hemani:2018ab}
Hemani, G., Zheng, J., Elsworth, B., Wade, K.~H., Haberland, V., Baird, D.,
  Laurin, C., Burgess, S., Bowden, J., Langdon, R., Tan, V.~Y., Yarmolinsky,
  J., Shihab, H.~A., Timpson, N.~J., Evans, D.~M., Relton, C., Martin, R.~M.,
  Davey~Smith, G., Gaunt, T.~R., Haycock, P.~C., and Loos, R. (2018).
\newblock The mr-base platform supports systematic causal inference across the
  human phenome.
\newblock {\em eLife}, 7:e34408.

\bibitem[Holmes et~al., 2017]{Holmes:2017}
Holmes, M.~V., Ala-Korpela, M., and Smith, G.~D. (2017).
\newblock Mendelian randomization in cardiometabolic disease: challenges in
  evaluating causality.
\newblock {\em Nature Reviews Cardiology}, 14(10):577--590.

\bibitem[Kang et~al., 2016]{Kang16}
Kang, H., Zhang, A., Cai, T.~T., and Small, D.~S. (2016).
\newblock Instrumental variables estimation with some invalid instruments and
  its application to mendelian randomization.
\newblock {\em Journal of the American Statistical Association},
  111(513):132--144.

\bibitem[Koles{\'a}r et~al., 2015]{Kole15}
Koles{\'a}r, M., Chetty, R., Friedman, J., Glaeser, E., and Imbens, G.~W.
  (2015).
\newblock Identification and inference with many invalid instruments.
\newblock {\em Journal of Business \& Economic Statistics}, 33(4):474--484.

\bibitem[Markozannes et~al., 2022]{markozannes2022systematic}
Markozannes, G., Kanellopoulou, A., Dimopoulou, O., Kosmidis, D., Zhang, X.,
  Wang, L., Theodoratou, E., Gill, D., Burgess, S., and Tsilidis, K.~K. (2022).
\newblock Systematic review of mendelian randomization studies on risk of
  cancer.
\newblock {\em BMC medicine}, 20(1):1--22.

\bibitem[Morrison et~al., 2020]{morrison2020ng}
Morrison, J., Knoblauch, N., Marcus, J.~H., Stephens, M., and He, X. (2020).
\newblock Mendelian randomization accounting for correlated and uncorrelated
  pleiotropic effects using genome-wide summary statistics.
\newblock {\em Nature Genetics}, 52(7):740–747.

\bibitem[Nieman et~al., 2012]{nieman2012psychosocial}
Nieman, P., Leblanc, C.~M., Society, C.~P., Living, H.~A., and Committee, S.~M.
  (2012).
\newblock Psychosocial aspects of child and adolescent obesity.
\newblock {\em Paediatrics \& child health}, 17(4):205--206.

\bibitem[Pingault et~al., 2018]{Pingault:2018aa}
Pingault, J.-B., O'Reilly, P.~F., Schoeler, T., Ploubidis, G.~B., Rijsdijk, F.,
  and Dudbridge, F. (2018).
\newblock Using genetic data to strengthen causal inference in observational
  research.
\newblock {\em Nature Reviews Genetics}, 19(9):566--580.

\bibitem[Qi and Chatterjee, 2019]{qi2019mendelian}
Qi, G. and Chatterjee, N. (2019).
\newblock Mendelian randomization analysis using mixture models for robust and
  efficient estimation of causal effects.
\newblock {\em Nature communications}, 10(1):1--10.

\bibitem[Richmond et~al., 2017]{richmond2017investigating}
Richmond, R., Wade, K., Corbin, L., Bowden, J., Hemani, G., Timpson, N., and
  Smith, G.~D. (2017).
\newblock Investigating the role of insulin in increased adiposity:
  Bi-directional mendelian randomization study.
\newblock {\em bioRxiv}, page 155739.

\bibitem[Rosenbaum, 2008]{rosenbaum2008testing}
Rosenbaum, P.~R. (2008).
\newblock Testing hypotheses in order.
\newblock {\em Biometrika}, 95(1):248--252.

\bibitem[Sanderson et~al., 2022]{sanderson2022mendelian}
Sanderson, E., Glymour, M.~M., Holmes, M.~V., Kang, H., Morrison, J.,
  Munaf{\`o}, M.~R., Palmer, T., Schooling, C.~M., Wallace, C., Zhao, Q.,
  et~al. (2022).
\newblock Mendelian randomization.
\newblock {\em Nature Reviews Methods Primers}, 2(1):1--21.

\bibitem[Sun et~al., 2021]{Sun2019}
Sun, B., Cui, Y., and Tchetgen, E.~T. (2021).
\newblock Selective machine learning of the average treatment effect with an
  invalid instrumental variable.
\newblock arXiv:1907.11882.

\bibitem[Tchetgen~Tchetgen et~al., 2021]{Tchetgen2019_GENIUS}
Tchetgen~Tchetgen, E.~J., Sun, B., and Walter, S. (2021).
\newblock The {GENIUS} approach to robust mendelian randomization inference.
\newblock {\em Statistical Science}, in press.

\bibitem[Verbanck et~al., 2018]{Verbanck:2018aa}
Verbanck, M., Chen, C.-Y., Neale, B., and Do, R. (2018).
\newblock Detection of widespread horizontal pleiotropy in causal relationships
  inferred from mendelian randomization between complex traits and diseases.
\newblock {\em Nature Genetics}, 50(5):693--698.

\bibitem[Windmeijer et~al., 2019]{Wind19}
Windmeijer, F., Farbmacher, H., Davies, N., and Davey~Smith, G. (2019).
\newblock On the use of the lasso for instrumental variables estimation with
  some invalid instruments.
\newblock {\em Journal of the American Statistical Association},
  114(527):1339--1350.

\bibitem[Windmeijer et~al., 2021]{Wind21}
Windmeijer, F., Liang, X., Hartwig, F.~P., and Bowden, J. (2021).
\newblock The confidence interval method for selecting valid instrumental
  variables.
\newblock {\em Journal of the Royal Statistical Society: Series B (Statistical
  Methodology)}.

\bibitem[Xue and Pan, 2020]{xue2020inferring}
Xue, H. and Pan, W. (2020).
\newblock Inferring causal direction between two traits in the presence of
  horizontal pleiotropy with gwas summary data.
\newblock {\em PLoS genetics}, 16(11):e1009105.

\bibitem[Ye et~al., 2021a]{ye2021genius}
Ye, T., Liu, Z., Sun, B., and Tchetgen, E.~T. (2021a).
\newblock Genius-mawii: For robust mendelian randomization with many weak
  invalid instruments.
\newblock {\em arXiv preprint arXiv:2107.06238}.

\bibitem[Ye et~al., 2021b]{ye2021debiased}
Ye, T., Shao, J., and Kang, H. (2021b).
\newblock Debiased inverse-variance weighted estimator in two-sample
  summary-data mendelian randomization.
\newblock {\em The Annals of Statistics}, 49(4):2079--2100.

\bibitem[Zhang et~al., 2018]{zhang2018estimation}
Zhang, Y., Qi, G., Park, J.-H., and Chatterjee, N. (2018).
\newblock Estimation of complex effect-size distributions using summary-level
  statistics from genome-wide association studies across 32 complex traits.
\newblock {\em Nature genetics}, 50(9):1318--1326.

\bibitem[Zhao et~al., 2019]{Zhao:2019aa}
Zhao, Q., Chen, Y., Wang, J., and Small, D.~S. (2019).
\newblock Powerful three-sample genome-wide design and robust statistical
  inference in summary-data mendelian randomization.
\newblock {\em International Journal of Epidemiology}, 48(5):1478--1492.

\bibitem[Zhao et~al., 2020]{zhao2018statistical}
Zhao, Q., Wang, J., Hemani, G., Bowden, J., and Small, D.~S. (2020).
\newblock Statistical inference in two-sample summary-data mendelian
  randomization using robust adjusted profile score.
\newblock {\em Annals of Statistics}, 48(3):1742--1769.

\end{thebibliography}

\end{document}